\documentclass{article}

\usepackage{arxiv}

\usepackage[utf8]{inputenc} % allow utf-8 input
\usepackage[T1]{fontenc}    % use 8-bit T1 fonts
\usepackage[hidelinks]{hyperref}       % hyperlinks
\usepackage{url}            % simple URL typesetting
\usepackage{booktabs}       % professional-quality tables
\usepackage{amsfonts}       % blackboard math symbols
\usepackage{nicefrac}       % compact symbols for 1/2, etc.
\usepackage{microtype}      % microtypography
\usepackage{lipsum}		% Can be removed after putting your text content
\usepackage[dvipdfmx]{graphicx}
\usepackage{natbib}
\usepackage{doi}

\usepackage{bm}
\usepackage{amsmath}
\usepackage{mathtools}
\usepackage{enumerate}
\usepackage{color}
\usepackage{algorithm}
\usepackage{algorithmic}

\newtheorem{Cor}{Corollary}

\newtheorem{Lem}{Lemma}
\newtheorem{Pro}{Proposition}
\newtheorem{Thm}{Theorem}

\newcommand{\df}{\mathop{\rm df}}
\newcommand{\diag}{\mathop{\rm diag}}
\newcommand{\rank}{\mathop{\rm rank}}

\newcommand{\tr}{\mathop{\rm tr}}
\newcommand{\vect}{\mathop{\rm vec}}

\newcommand{\Cov}{\mathop{\rm Cov}\nolimits}
\newcommand{\E}{\mathop{\rm E}\nolimits}

\newcommand{\MSE}{\mathop{\rm MSE}\nolimits}
\newcommand{\PRSS}{\mathop{\rm PRSS}\nolimits}

\newcommand{\RSS}{\mathop{\rm RSS}\nolimits}
\newcommand{\Var}{\mathop{\rm Var}\nolimits}

\renewcommand{\vec}[1]{{\bm{#1}}}
\def\0{\vec{0}}
\def\1{\vec{1}}

\def\va{\vec{a}}
\def\vb{\vec{b}}

\def\ve{\vec{e}}

\def\vh{\vec{h}}

\def\vr{\vec{r}}
\def\vs{\vec{s}}

\def\vu{\vec{u}}
\def\vv{\vec{v}}

\def\vx{\vec{x}}

\def\vz{\vec{z}}

\def\vA{\vec{A}}

\def\vF{\vec{F}}

\def\vH{\vec{H}}
\def\vI{\vec{I}}

\def\vO{\vec{O}}

\def\vQ{\vec{Q}}
\def\vR{\vec{R}}
\def\vS{\vec{S}}

\def\vU{\vec{U}}
\def\vV{\vec{V}}

\def\vX{\vec{X}}
\def\vY{\vec{Y}}
\def\vZ{\vec{Z}}

\def\cA{\mathcal{A}}
\def\cB{\mathcal{B}}

\def\cR{\mathcal{R}}

\def\bR{\mathbb{R}}

\def\vbeta{\vec{\beta}}
\def\vgamma{\vec{\gamma}}

\def\vtheta{\vec{\theta}}
\def\vlambda{\vec{\lambda}}
\def\vmu{\vec{\mu}}

\def\vxi{\vec{\xi}}

\def\vEpsilon{\vec{\mathcal{E}}}
\def\vTheta{\vec{\Theta}}
\def\vXi{\vec{\Xi}}
\def\vSigma{\vec{\Sigma}}

\def\vPsi{\vec{\Psi}}
\def\vOmega{\vec{\Omega}}

\title{Hierarchical Overlapping Group Lasso\\ for GMANOVA Model}

\author{ 
  \href{https://orcid.org/0000-0001-8727-3631}{\includegraphics[scale=0.06]{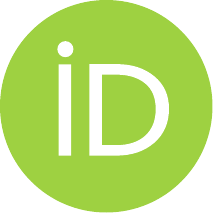}\hspace{1mm}Mineaki Ohishi}\thanks{
    Corresponding author 
  } \\
  Center for Data-driven Science and Artificial Intelligence \\
  Tohoku University\\
  Sendai, Japan \\
  \texttt{mineaki.ohishi.a4@tohoku.ac.jp} \\
  \And
  Isamu Nagai \\
  Faculty of Liberal Arts and Science \\
  Chukyo University \\
  Nagoya, Japan 
  \And
  Ryoya Oda \\
  Graduate School of Advanced Science and Engineering\\
  Hiroshima University\\
  Higashi-Hiroshima, Japan 
  \And
  Hirokazu Yanagihara \\
  Osaka Central Advanced Mathematical Institute \\
  Osaka Metropolitan University\\
  Osaka, Japan \\
  Research \& Development Center \\
  Osaka Medical and Pharmaceutical University \\
  Osaka, Japan
}

\renewcommand{\headeright}{}
\renewcommand{\undertitle}{}
\renewcommand{\shorttitle}{HOGL for GMANOVA}

\begin{document}
\maketitle

%%%%%%%%%%%%%%%%%%%%%%%%%%%%%%
%     Abstract
%%%%%%%%%%%%%%%%%%%%%%%%%%%%%%
\begin{abstract}
  This paper deals with the GMANOVA model with a matrix of polynomial basis functions as a within-individual design matrix.
  The model involves two model selection problems: the selection of explanatory variables and the selection of the degrees of the polynomials.
  The two problems can be uniformly addressed by hierarchically incorporating zeros into the vectors of regression coefficients.
  Based on this idea, we propose hierarchical overlapping group Lasso (HOGL) to perform the variable and degree selections simultaneously.
  Importantly, when using a polynomial basis, fitting a high-degree polynomial often causes problems in model selection.
  In the approach proposed here, these problems are handled by using a matrix of orthonormal basis functions obtained by transforming the matrix of polynomial basis functions.
  Algorithms are developed with optimality and convergence to optimize the method. The performance of the proposed method is evaluated using numerical simulation.
\end{abstract}

% keywords can be removed

%%%%%%%%%%%%%%%%%%%%%%%%%%%%%%
%     Sec1 
%%%%%%%%%%%%%%%%%%%%%%%%%%%%%%
\section{Introduction}

Suppose we have observations at $p$ common time points for $n\ (> p)$ individuals and define $\vY$ as an $n \times p$ matrix of the observations.
Let $\vA$ and $\vX$ be an $n \times k\ (n > k)$ between-individual design matrix and a $p \times q\ (p \ge q)$ within-individual design matrix, respectively.
For these matrices, we consider the following GMANOVA model \citep{PotthoffRoy1964}:
\begin{align}
\label{model}
  \vY
    = \1_n \vmu' + \vA \vTheta \vX' + \vEpsilon, \quad
  \E [\vEpsilon] = \vO_{n, p},\ 
  \Cov [\vect (\vEpsilon)] = \vSigma \otimes \vI_n,
\end{align}
where $\1_m$ is an $m$-dimensional vector of ones, $\vmu = (\mu_1, \ldots, \mu_p)'$ is a $p$-dimensional vector of location parameters, $\vTheta = (\vtheta_1, \ldots, \vtheta_k)'$ is a $k \times q$ matrix of regression coefficients, $\vO_{r, s}$ is an $r \times s$ matrix of zeros, $\vSigma$ is a $p \times p$ covariance matrix, $\vect (\cdot)$ represents a vec operator, and $\otimes$ represents the Kronecker product.
Furthermore, suppose that $\vA$ is centralized (i.e., $\vA' \1_n = \0_k$) and $\vA$ and $\vX$ are column full rank (i.e., $\rank (\vA) = k$, $\rank (\vX) = q$) and standardized such that a norm of each column vector is one, where $\0_m$ is an $m$-dimensional vector of zeros.
Note that the GMANOVA model is a generalized version of a multivariate linear regression model [e.g., \citealp{Srivastava2002}, Chap.~9; \citealp{Timm2002}, Chap.~4].
When $p = q$ and $\vX = \vI_p$, the GMANOVA model reduces to a multivariate linear regression model.
Furthermore, when $p=1$, the GMANOVA model reduces to a univariate linear regression model.

The GMANOVA model is also referred to as the growth curve model [e.g., \citealp{vonRosen1991}; \citealp{KshirsagarSmith1995}] and has aspects of a varying coefficient model \citep{SatohYanagihara2010}.
Let $t_1, \ldots, t_p$ be common time points of $n$ observations and define $\vX = (\vx (t_1), \ldots, \vx (t_p))'$, where $\vx (t)$ is a $q$-dimensional vector of basis functions at time $t$.
Then, $y_{i, j}$, the $(i, j)$th element of $\vY$, can be expressed as the following varying coefficient model:
\begin{align*}
  y_{i, j}
    = \mu_j + \sum_{\ell = 1}^k a_{i, \ell} \beta_\ell (t_j) + \varepsilon_{i, j}, \quad
  \beta_\ell (t)
    = \vtheta_\ell' \vx (t),
\end{align*}
where $a_{i, j}$ and $\varepsilon_{i, j}$ are the $(i, j)$th elements of $\vA$ and $\vEpsilon$, respectively, and $a_{i, \ell}$ is the $\ell$th explanatory variable of the $i$th individual.
This paper adopts the following vector of polynomial basis functions:
\begin{align*}
  \vx (t)
    = \left( c_{q-1} t^{q-1}, \ldots, c_1 t, p^{-1/2} \right)', \quad
  c_j = \left( \sum_{\ell=1}^p t_\ell^{2 j} \right)^{-1/2},
\end{align*}
where $c_j$ is a constant standardizing column vectors of $\vX$.
This implies that the time trend for each explanatory variable is modeled by the following polynomial of degree $(q-1)$:
\begin{align}
\label{beta}
  \beta_\ell (t)
    = \theta_{\ell, 1} c_{q-1} t^{q-1} + \cdots + \theta_{\ell, q-1} c_1 t + \theta_{\ell, q} p^{-1/2},
\end{align}
where $\theta_{\ell, j}$ is the $j$th element of $\vtheta_\ell$.
We consider model selection for the GMANOVA model \eqref{model} with the above settings.

As an example of model selection for a GMANOVA model, \cite{FujikoshiRao1991} dealt with selection of the explanatory variables, while \cite{Satoh-1997} and \cite{Enomoto-2015} dealt with selection of the degrees of the polynomial basis functions.
Although they are typical approaches that select the best model from candidates based on a model selection criterion, sparse estimation methods offer a different approach and are popular for model selection.
Sparse estimation methods can perform parameter estimation and model selection simultaneously by shrinking the parameters towards zero and allowing some estimates to be exactly zero.
These methods are known to be useful for models with a large number of parameters.
Following the proposal of Lasso by \cite{Tibshirani1996}, various specific methods have been offered, e.g., fused Lasso \citep{Tibshirani-2005}, group Lasso \citep{YuanLin2006}, and sparse group Lasso \citep{Simon-2013}.

This paper proposes hierarchical overlapping group Lasso (HOGL) to perform the selection of the explanatory variables and the selection of the degrees of the polynomial basis functions simultaneously.
Regarding variable selection, for multivariate models such as the GMANOVA model, the general approach is to select variables that affect at least one response variable [e.g., \citealp{Obozinski-2008}; \citealp{YanagiharaOda2021}].
In our model, variable selection means the selection of column vectors of $\vA$ where $\vtheta_\ell = \0_q$ implies that the $\ell$th column vector (i.e., the $\ell$th explanatory variable) is removed from the model.
Hence, variable selection can be performed by group Lasso for row vectors of $\vTheta$, which can be implemented by penalized estimation with $\sum_{\ell=1}^k \| \vtheta_\ell \|$.
On the other hand, regarding degree selection, by setting, for example, all the coefficients of the fourth degree and higher to zero, i.e., $\theta_{\ell, 1} = \cdots = \theta_{\ell, q-4} = 0$, $\beta_\ell (t)$ reduces to a cubic polynomial.
Hence, degree selection can be performed by hierarchically applying group Lasso to elements of $\vtheta_\ell$ beginning with higher-order coefficients, and we can consider employing $\sum_{j=1}^q \| (\vtheta_\ell)_{(j)} \|$ as a penalty term, where $\vgamma_{(j)}$ is a sub-vector of an $m$-dimensional vector $\vgamma = (\gamma_1, \ldots, \gamma_m)'$ defined by 
\begin{align}
\label{subvec}
  \vgamma_{(j)} = (\gamma_1, \ldots, \gamma_j)' \quad
  (j \in \{ 1, \ldots, m \}), 
\end{align}
and $\vgamma_{(m)} = \vgamma$ holds.
Based on the above, we define the HOGL penalty as
\begin{align}
\label{HOGLP}
  \Omega (\vTheta \mid \delta)
    &= \delta \sum_{\ell=1}^k \sum_{j=1}^q w_{\ell, j}^{(0)} \| (\vtheta_\ell)_{(j)} \|
           + (1 - \delta) \sum_{\ell=1}^k w_{\ell, q}^{(0)} \| \vtheta_\ell \| 
    = \sum_{\ell=1}^k \sum_{j=1}^q w_{\ell, j} (\delta) \| (\vtheta_\ell)_{(j)} \|, \\
  w_{\ell, j} (\delta) &= \begin{dcases}
      \delta w_{\ell, j}^{(0)} &(j = 1, \ldots, q-1) \\
      w_{\ell, q}^{(0)} &(j = q)
    \end{dcases}, \quad
  \delta \in [0, 1], \nonumber
\end{align}
where $\delta$ is a tuning parameter adjusting the balance of the penalties for the variable and degree selections and $w_{\ell, j}^{(0)}\ (>0)$ is a penalty weight.
With the HOGL penalty, estimations of $\vmu$ and $\vTheta$ and the variable and degree selections are simultaneously performed by minimizing the following function:
\begin{align}
\label{obje}
  \dfrac{1}{2} \tr \left\{ (\vY - \1_n \vmu' - \vA \vTheta \vX')' (\vY - \1_n \vmu' - \vA \vTheta \vX') \vS^{-1} \right\}
    + \lambda \Omega (\vTheta \mid \delta),
\end{align}
where $\lambda\ (\ge 0)$ is a tuning parameter adjusting the strength of the penalty against model fitting and $\vS$ is an unbiased estimator of $\vSigma$ defined as
\begin{align*}
  \vS = \dfrac{\vY' \{ \vI_n - \1_n \1_n' / n - \vA (\vA' \vA)^{-1} \vA' \} \vY}{n - k - 1}.
\end{align*}

As a model selection method based on sparse estimation for the GMANOVA model, weighted least squares estimation with group SCAD penalty (we call it wSCAD here) was proposed by \cite{Hu-2014}.
wSCAD performs variable selection and degree selection simultaneously based on SCAD \citep{FanLi2001} and is guaranteed to have oracle properties \citep{FanLi2001} under specific conditions.
One of the main differences between HOGL and wSCAD is their approach to degree selection.
As described above, we can express $\beta_\ell (t)$ as a cubic polynomial by setting all the coefficients of fourth degree and higher to zero.
Whereas HOGL can set the coefficients to zero simultaneously, wSCAD sets the coefficients to zero individually.
From this, we can expect that HOGL mitigates the risks of both missing the truly zero coefficients and mistakenly shrinking non-zero coefficients to zero.

To implement HOGL, an algorithm for minimizing \eqref{obje} is important.
Since $\vA$ is centralized, the estimator of $\vmu$ is given by $\hat{\vmu} = \vY' \1_n / n$.
Hence, it is sufficient to minimize \eqref{obje} with respect to $\vTheta$.
To achieve this, we propose an algorithm with optimality and convergence based on the MM philosophy \citep{HunterLange2004}.
Importantly, we construct a practical algorithm by deriving the update equation of a solution in closed form.
Recognizing that using a polynomial basis can result in model unstableness in the case of high-degree polynomials, which, in turn, can lead to model misspecification in the model selection process,
we transform the matrix of polynomial basis functions $\vX$ to a matrix of orthonormal basis functions.
To choose the number of orthonormal basis functions for selecting the degrees of the original polynomial basis functions, we apply HOGL.
To use HOGL with the matrix of orthonormal basis functions, we construct an algorithm based on the block-wise coordinate descent method.
Moreover, we discuss an extension of HOGL.
As shown in \eqref{HOGLP}, a polynomial basis requires hierarchical overlapping by adding parameters one by one to select the degree of the polynomial.
However, this is not necessarily the case for all basis types. For example, a Fourier basis has pairs $\sin$ and $\cos$, and their selection requires hierarchical overlapping by adding parameters two by two.
To select various types of basis functions, we extend HOGL to a flexible version of hierarchical overlapping.

The remainder of the paper is organized as follows. 
In Section 2, we establish the foundation of our study.
In Section 3, we describe the study's main results.
By deriving the update equation of a solution in closed form, we construct an algorithm to optimize HOGL.
We also describe the transformation of the matrix of basis functions and our extension of HOGL.
In Section 4, we numerically evaluate the performance of the proposed HOGL method. Section 5 concludes the paper.
Technical details are provided in the Appendices.

%%%%%%%%%%%%%%%%%%%%%%%%%%%%%%
%     Sec2
%%%%%%%%%%%%%%%%%%%%%%%%%%%%%%
\section{Preliminaries}

Since $\vA$ is centralized, the first term of \eqref{obje} can be separated with respect to $\vmu$ and $\vTheta$ as
\begin{align*}
  \dfrac{1}{2} \tr \left\{ (n \vmu \vmu' - \vY' \1_n \vmu' - \vmu \1_n' \vY) \vS^{-1} \right\}
    + \dfrac{1}{2} \tr \left\{ (\vY - \vA \vTheta \vX')' (\vY - \vA \vTheta \vX') \vS^{-1} \right\}.
\end{align*}
Let $\vU = \vY \vS^{-1/2}$ and $\vV = \vS^{-1/2} \vX$. 
Then, the second term of the above equation can be expressed as 
\begin{align}
\label{RSS} 
  \RSS (\vtheta)
    &= \dfrac{1}{2} \| \vu - \vZ \vtheta \|^2 
    = \dfrac{1}{2} \tr \left\{ (\vU - \vA \vTheta \vV')' (\vU - \vA \vTheta \vV')  \right\},
\end{align}
where $\vu = \vect (\vU')$, $\vtheta = \vect (\vTheta') = (\vtheta_1', \ldots, \vtheta_k')'$, and $\vZ = \vA \otimes \vV$.
Hence, the estimation problem of $\vTheta$ based on minimizing \eqref{obje} is equal to the minimization problem of the following penalized residual sum of squares (PRSS).
\begin{align}
\label{PRSS}
  \PRSS (\vtheta \mid \delta, \lambda)
    &= \RSS (\vtheta) 
           + \lambda \Omega (\vTheta \mid \delta).
\end{align}

To construct an algorithm to minimize \eqref{PRSS}, we define a surrogate function of $\RSS$.
Let $L\ (> 0)$ be the maximum eigenvalue of $\vZ' \vZ$ and $\vr (\vtheta)$ be a gradient vector of $\RSS (\vtheta)$ as
\begin{align*}
  \vr (\vtheta)
    = \left( \vr_1 (\vtheta)', \ldots, \vr_k (\vtheta)' \right)'
    = \dfrac{\partial \RSS}{\partial \vtheta} (\vtheta)
    = \vZ' \vZ \vtheta - \vZ' \vu.
\end{align*}
Using them, we define
\begin{align*}
  \RSS^+ \left( \vtheta \mid \hat{\vtheta} \right)
    &= L \sum_{\ell=1}^k \left[
           \dfrac{1}{2} \| \vtheta_\ell \|^2 - \left\{ \hat{\vtheta}_\ell - \dfrac{\vr_\ell (\hat{\vtheta})}{L} \right\}' \vtheta_\ell
         \right] + \RSS (\hat{\vtheta}) - \vr (\hat{\vtheta})' \hat{\vtheta} + \dfrac{L}{2} \| \hat{\vtheta} \|^2, \\
  \hat{\vtheta} 
    &= \left( \hat{\vtheta}_1', \ldots, \hat{\vtheta}_k' \right)' \in \bR^{kq}.
\end{align*}
Then, $\RSS$ and $\RSS^+$ have the following relationships:
\begin{align}
\label{RSS+}
  \RSS^+ \left( \hat{\vtheta} \mid \hat{\vtheta} \right) = \RSS (\hat{\vtheta}), \quad
  \RSS^+ \left( \vtheta \mid \hat{\vtheta} \right) \ge \RSS (\vtheta).
\end{align}
These can be obtained by Taylor expansion of $\RSS (\vtheta)$ around $\vtheta = \hat{\vtheta}$ and $\RSS^+$ is a surrogate function of $\RSS$.

To solve the optimization problem for HOGL, the essential task is to minimize the following function:
\begin{align}
\label{f}
  f (\vgamma)
    &= \dfrac{1}{2} \| \vgamma \|^2 - \vb' \vgamma 
          + \sum_{j=1}^q \lambda_j \| \vgamma_{(j)} \|, \quad
  \vgamma
    = (\gamma_1, \ldots, \gamma_q)',
\end{align}
where $\vb = (b_1, \ldots, b_q)'$ is a $q$-dimensional vector of constants, $\lambda_j$ is a non-negative constant, and $\vgamma_{(j)}$ is a sub-vector of $\vgamma$ given by \eqref{subvec}.
A key to minimizing $f (\vgamma)$ is $d_{\alpha, j}$ defined by
\begin{align}
\label{d}
 d_{\alpha, j} &= \begin{dcases}
   0 &(j < \alpha) \\
   \left( |b_j| - \lambda_j \right)_+
     & (j = \alpha) \\
   \left( \sqrt{d_{\alpha, j-1}^2 + b_j^2} - \lambda_j \right)_+
     & (\alpha < j)
   \end{dcases} \quad (\alpha, j \in \{ 1, \ldots, q \}),
\end{align}
where $(x)_+ = \max \{ x, 0 \}$.
Regarding $d_{\alpha, j}$, we have the following proposition (the proof is given in Appendix~\ref{ap propd}).
\begin{Pro}
\label{propd}
  The $d_{\alpha, j}$ in \eqref{d} satisfies $d_{\alpha, j} \ge d_{\alpha+1, j}$.
\end{Pro}
By putting $d_{\alpha, j}$ into the $(\alpha, j)$th element, we have the following upper triangular matrix.
\begin{align}
\label{D}
\begin{pmatrix}
  \vspace{0.4cm}
  (|b_1| - \lambda_1)_+ & \left( \sqrt{d_{1, 1}^2 + b_2^2} - \lambda_2 \right)_+ & \cdots & \left( \sqrt{d_{1, q-1}^2 + b_q^2} - \lambda_q \right)_+ \\
  \vspace{0.4cm}
  0 & (|b_2| - \lambda_2)_+ & \ddots & \vdots \\
  \vspace{0.4cm}
  \vdots & \ddots & \quad \ddots \quad & \left( \sqrt{d_{q-1, q-1}^2 + b_q^2} - \lambda_q \right)_+ \\
  0 & \cdots & 0 & (|b_q| - \lambda_q)_+ 
\end{pmatrix}.
\end{align}
Proposition~\ref{propd} means the elements of the above matrix are monotonically decreasing in each column.

%%%%%%%%%%%%%%%%%%%%%%%%%%%%%%
%     Sec3
%%%%%%%%%%%%%%%%%%%%%%%%%%%%%%
\section{Main results}

The minimizer of $f (\vgamma)$ in \eqref{f} is given by the following theorem (the proof is given in Appendix~\ref{ap mainTh}).
\begin{Thm}
\label{th main}
  Define $\alpha_\ast$ as
  \begin{align*}
    \alpha_\ast &= \begin{dcases}
        \min \cA &(\cA \neq \emptyset) \\
        q + 1 &(\cA = \emptyset)
      \end{dcases}, \quad
    \cA = \left\{
        \alpha \in \{ 1, \ldots, q \} \mid \forall j \in \{ \alpha, \alpha + 1, \ldots, q \}, d_{\alpha, j} > 0
      \right\},
  \end{align*}
  where $d_{\alpha, j}$ is given by \eqref{d}.
  Let $\vgamma^\ast = (\gamma_1^\ast, \ldots, \gamma_q^\ast)'$ be the minimizer of $f (\vgamma)$, i.e., $\vgamma^\ast = \arg \min_{\vgamma \in \bR^q} f (\vgamma)$.
  Then, $\gamma_j^\ast$ is given by
  \begin{align*}
    \gamma_j^\ast = \begin{dcases}
        0 &(j < \alpha_\ast) \\
        b_j \prod_{\ell=j}^q \dfrac{d_{\alpha_\ast, \ell}}{d_{\alpha_\ast, \ell} + \lambda_\ell} 
          &(\alpha_\ast \le j)
      \end{dcases}.
  \end{align*}
\end{Thm}
Applying Theorem~\ref{th main} to obtain $\vgamma^\ast$ requires searching for $\alpha_\ast$.
We can obtain $\alpha_\ast$ directly by searching for a row for which all elements are positive in the upper triangle part of the matrix \eqref{D}. This requires the calculation of $O (q^2)$.
Fortunately, Proposition~\ref{propd} makes the search of $\alpha_\ast$ efficient.
Since Proposition~\ref{propd}, $d_{\alpha + 1, j} = d_{\alpha+2, j} = \cdots = d_{q, j} = 0$ holds when $d_{\alpha, j} = 0$.
This fact provides Algorithm~\ref{alg d2}.
\begin{algorithm}[h]
\caption{Efficient method for searching $\alpha_\ast$}
\label{alg d2}
  \begin{algorithmic}
    \STATE set $\alpha \gets 1$
    \FOR{$j = 1, \ldots, q$}
      \STATE calculate $d_{\alpha, j}$
      \IF{$d_{\alpha, j} = 0$}
        \STATE set $\alpha \gets j + 1$
      \ENDIF
    \ENDFOR
    \STATE define $\alpha_\ast = \alpha$
  \end{algorithmic}
\end{algorithm}
If $d_{\alpha, j}$ is positive, move to the next column in the same row ($d_{\alpha, j+1}$). 
If $d_{\alpha, j}$ is zero, since other elements in the same column are all zero, move to the diagonal element in the next column ($d_{j+1, j+1}$).
By starting from $d_{1, 1}$ and repeating the above procedure, we can obtain $\alpha_\ast$ with the calculation of $O (q)$.
With Theorem~\ref{th main} and Algorithm~\ref{alg d2}, we construct the algorithm to solve the optimization problem for HOGL.

%=============================
%     Sec3.1
%=============================
\subsection{Optimization of hierarchical overlapping group Lasso}
\label{sec HOGL}

To perform the selection of the explanatory variables and the degrees of the polynomial basis functions simultaneously for the GMANOVA model \eqref{model}, we estimate $\vtheta\ (= \vect (\vTheta'))$ based on minimizing $\PRSS (\vtheta \mid \delta, \lambda)$ in \eqref{PRSS}.
For this minimization, we construct an algorithm based on the MM philosophy.
Since \eqref{RSS+} and strict convexity of $\PRSS$, \cite{Razaviyayn-2013} gives the following theorem.
\begin{Thm}
\label{th MMconv}
  Let $\vtheta^{(0)}$ be an initial vector and consider the following update of a solution.
  \begin{align}
  \label{MMupdate}
    \vtheta^{(i)}
      &= \arg \min_{\vtheta} \PRSS^+ \left( \vtheta \mid \vtheta^{(i-1)}, \delta, \lambda \right) \quad (i = 1, 2, \ldots), \\
    \PRSS^+ \left( \vtheta \mid \hat{\vtheta}, \delta, \lambda \right)
      &= \RSS^+ \left( \vtheta \mid \hat{\vtheta} \right)
             + \lambda \Omega (\vTheta \mid \delta). \nonumber
  \end{align}
  Then, $\vtheta^{(i)}$ converges to the minimizer of $\PRSS (\vtheta \mid \delta, \lambda)$.
\end{Thm}
From the theorem, we can obtain the minimizer of $\PRSS (\vtheta \mid \delta, \lambda)$ by repeatedly solving the minimization problem of $\PRSS^+ (\vtheta \mid \hat{\vtheta}, \delta, \lambda)$.
Notice that $\PRSS^+ (\vtheta \mid \hat{\vtheta}, \delta, \lambda)$ is separable with respect to $\vtheta_\ell\ (\ell = 1, \ldots, k)$.
Hence, the minimization problem of $\PRSS^+ (\vtheta \mid \hat{\vtheta}, \delta, \lambda)$ reduces to that of the following function for each $\vtheta_\ell$:
\begin{align}
\label{theta ell}
  \dfrac{1}{2} \| \vtheta_\ell \|^2 - \left\{ \hat{\vtheta}_\ell - \dfrac{\vr_\ell (\hat{\vtheta})}{L} \right\}' \vtheta_\ell
    + \sum_{j=1}^q \lambda_{\ell, j} \| (\vtheta_\ell)_{(j)} \|, \quad
  \lambda_{\ell, j}
    = \dfrac{\lambda w_{\ell, j} (\delta)}{L}.
\end{align}
It is obvious that this function is essentially equal to \eqref{f}. 
Thus, we can obtain the minimizer in closed form by Theorem~\ref{th main} and Algorithm~\ref{alg d2}.
This is summarized as follows.
\begin{Cor}
\label{cor1}
  The minimizer of $\PRSS^+ (\vtheta \mid \hat{\vtheta}, \delta, \lambda)$ is given in closed form following Theorem~\ref{th main}.
\end{Cor}
The details of the algorithm for minimizing $\PRSS (\vtheta \mid \delta, \lambda)$ are summarized in Algorithm~\ref{alg main1}.
\begin{algorithm}[H]
\caption{MM algorithm for minimizing $\PRSS (\vtheta \mid \delta, \lambda)$}
\label{alg main1}
  \begin{algorithmic}
    \REQUIRE $\delta$, $\lambda$ and an initial vector $\vtheta^{(0)}$
    \STATE set $i \gets 0$
    \REPEAT
      \STATE set $i \gets i + 1$
      \STATE calculate $\vtheta^{(i)}$ by minimizing $\PRSS^+ (\vtheta \mid \vtheta^{(i-1)}, \delta, \lambda)$ which is minimized by the following procedure
      \FOR{$\ell = 1, \ldots, k$}
        \STATE calculate $\vtheta_\ell^{(i)}$ by applying Theorem~\ref{th main} with Algorithm~\ref{alg d2} to the minimization of \eqref{theta ell} with $\hat{\vtheta} = \vtheta^{(i-1)}$
      \ENDFOR
    \UNTIL{$\vtheta^{(i)}$ converges}
  \end{algorithmic}
\end{algorithm}
From Theorem~\ref{th MMconv} and Corollary~\ref{cor1}, it is guaranteed that the solution obtained by Algorithm~\ref{alg main1} converges to the optimal solution.
Note that although a for loop is used in Algorithm~\ref{alg main1} to minimize $\PRSS^+ (\vtheta \mid \hat{\vtheta}, \delta, \lambda)$, this procedure is parallelizable.

In the implementation of HOGL, since $\PRSS$ includes two tuning parameters, $\delta$ and $\lambda$, selection of the two parameters is required.
The search range of $\delta$ is $[0, 1]$, while that of $\lambda$ is all positive real values.
To limit the search range of $\lambda$, it is desirable to find $\lambda$ such that $\PRSS (\vtheta \mid \delta, \lambda)$ is minimized at $\vtheta = \0_{kq}$.
A sufficient condition for such $\lambda$ is given by the following theorem (the proof is given in Appendix \ref{ap thlam}).
\begin{Thm}
\label{th lam}
  Define $\lambda_{\max}$ as
  \begin{align*}
    \lambda_{\max} (\delta)
      = \max_{\ell \in \{ 1, \ldots, k \}, j \in \{ 1, \ldots, q \}} \dfrac{|\vu' \vz_{\ell j}|}{w_{\ell, j} (\delta)},
  \end{align*}
  where $\vz_{\ell j} = (a_{1, \ell} \vv_j', \ldots, a_{n, \ell} \vv_j')'$, $a_{i, \ell}$ is the $(i, \ell)$th element of $\vA$, and $\vv_j$ is the $j$th column vector of $\vV$.
  Then, we have
  \begin{align*}
    \lambda \ge \lambda_{\max} (\delta)
      \Longrightarrow \PRSS (\0_{kq} \mid \delta, \lambda) < \PRSS (\vtheta \mid \delta, \lambda) \quad 
      (\forall \vtheta \in \bR^{kq} \backslash \{ \0_{kq} \}).
  \end{align*}
\end{Thm}
From the theorem, we should select the optimal $\lambda$ in $[0, \lambda_{\max}]$.

%=============================
%     Sec3.2
%=============================
\subsection{Transformation of the matrix of basis functions}
\label{sec trans}

In the previous section, we proposed HOGL to select the explanatory variables and the degrees of the polynomial basis functions simultaneously.
However, we need to consider the potential problems that can be caused by the degrees of the polynomials.
When $q$, the dimension of the polynomial basis, increases, $t_j^{q-1}$ diverges to infinity if $|t_j| > 1$ and $t_j^{q-1}$ converges to zero if $|t_j| < 1$.
Hence, fitting high-degree polynomials makes the model unstable and renders model selection difficult.
To address this problem, we consider transformation of the matrix of basis functions.

Let $\vv_1, \ldots, \vv_q$ be column vectors of $\vV$ in \eqref{RSS}.
Gram-Schmidt orthogonalization provides orthonormal basis vectors $\vh_q, \vh_{q-1}, \ldots, \vh_1$ from $\vv_q, \vv_{q-1}, \ldots, \vv_1$ and define $\vH = (\vh_1, \ldots, \vh_q)$.
Then, we have $\vV = \vH \vQ$, where $\vQ$ is a $q \times q$ lower triangular matrix (the details are given in Appendix~\ref{ap QR}).
This decomposition implies $\vA \vTheta \vV' = \vA \vXi \vH' \ (\vXi = \vTheta \vQ')$ and $\RSS$ in \eqref{RSS} is rewritten as
\begin{align*}
  \RSS^\dagger (\vxi)
    &= \dfrac{1}{2} \| \vu - \vZ^\dagger \vxi \|^2,
\end{align*}
where $\vxi = \vect (\vXi')$ and $\vZ^\dagger = \vA \otimes \vH$.
Hence, we consider the estimation of $\vXi$ instead of $\vTheta$.
This transformation means that a model with a polynomial basis is transformed to one with an orthonormal basis.
Then, to select degrees of polynomial basis functions in the original model, we need to select the number of orthonormal basis vectors.
Fortunately, HOGL can be applied to the selection of the orthonormal basis vectors, and hence we can estimate $\vXi$ by minimizing the following PRSS:
\begin{align*}
  \PRSS^\dagger (\vxi \mid \delta, \lambda)
    &= \RSS^\dagger (\vxi)
        + \lambda \sum_{\ell=1}^k \sum_{j=1}^q w_{\ell, j} (\delta) \| (\vxi_\ell)_{(j)} \|, \quad
  \vXi = (\vxi_1, \ldots, \vxi_k)'.
\end{align*}
By applying HOGL for $\vxi_\ell$, we can perform the selections of the explanatory variables and the number of orthonormal basis vectors simultaneously.
Assume that we have the estimator $\hat{\vXi}$ of $\vXi$ by minimizing $\PRSS^\dagger (\vxi \mid \delta, \lambda)$.
Then, the estimator of $\vTheta$ is given by $\hat{\vTheta} = \hat{\vXi} (\vQ')^{-1}$ and the hierarchical structure in $\hat{\vXi}$ is inherited by $\hat{\vTheta}$ because $(\vQ')^{-1}$ is an upper triangular matrix.
Hence, we can select degrees of the original polynomials via the matrix of orthonormal basis.

Although Algorithm~\ref{alg main1} can be applied to the minimization of $\PRSS^\dagger (\vxi \mid \delta, \lambda)$ because it is essentially equal to $\PRSS (\vtheta \mid \delta, \lambda)$, we can directly apply the block-wise coordinate descent method to the minimization problem by the orthogonality of $\vH$.
We divide $\vZ^\dagger$ into the blocks as
\begin{align*}
  \vZ^\dagger
    = \left( \vZ_1^\dagger, \ldots, \vZ_k^\dagger \right), \quad
  \vZ_\ell^\dagger
    = \va_\ell \otimes \vH,
\end{align*}
where $\va_\ell$ is the $\ell$th column vector of $\vA$.
Since $\| \va_\ell \| = 1$ and $\vH' \vH = \vI_q$, we can separate $\RSS^\dagger (\vxi)$ with respect to $\vxi_\ell$ as
\begin{align*}
  \RSS^\dagger (\vxi)
    &= \dfrac{1}{2} \| \tilde{\vu}_\ell - \vZ_\ell^\dagger \vxi_\ell \|^2
    = \dfrac{1}{2} \| \vxi_\ell \|^2 - \tilde{\vu}_\ell' \vZ_\ell^\dagger \vxi_\ell + \dfrac{1}{2} \| \tilde{\vu}_\ell \|^2, \quad
  \tilde{\vu}_\ell
    = \vu - \sum_{j \neq \ell}^k \vZ_j^\dagger \vxi_j.
\end{align*}
Hence, $\vxi_\ell$ is updated by minimizing the following function:
\begin{align}
\label{xi ell}
  \dfrac{1}{2} \| \vxi_\ell \|^2 - \tilde{\vu}_\ell' \vZ_\ell^\dagger \vxi_\ell
    + \sum_{j=1}^q \lambda_{\ell, j}^\dagger \| (\vxi_\ell)_{(j)} \|, \quad
  \lambda_{\ell, j}^\dagger
    = \lambda w_{\ell, j} (\delta).
\end{align}
This function is essentially equal to \eqref{f}, and hence its minimizer is obtained in closed form by Theorem~\ref{th main}.
The details of the algorithm for minimizing $\PRSS^\dagger (\vxi \mid \delta, \lambda)$ are summarized in Algorithm~\ref{alg main2}.
\begin{algorithm}[h]
\caption{Blockwise coordinate descent method for minimizing $\PRSS^\dagger (\vxi \mid \delta, \lambda)$}
\label{alg main2}
  \begin{algorithmic}
    \REQUIRE $\delta$, $\lambda$ and an initial vector $\vxi^{(0)}$
    \STATE set $i \gets 0$
    \REPEAT
      \STATE set $i \gets i + 1$
      \STATE calculate $\vxi^{(i)} = ({\vxi_1^{(i)}}', \ldots, {\vxi_k^{(i)}}')'$ by the following procedure
      \FOR{$\ell = 1, \ldots, k$}
        \STATE calculate $\vxi_\ell^{(i)}$ by applying Theorem~\ref{th main} with Algorithm~\ref{alg d2} to the minimization of \eqref{xi ell} with $\vxi_j = \vxi_j^{(i)}\ (j < \ell)$ and $\vxi_j = \vxi_j^{(i-1)}\ (j > \ell)$
      \ENDFOR
    \UNTIL{$\vxi^{(i)}$ converges}
  \end{algorithmic}
\end{algorithm}
Notice that the penalty term of $\PRSS^\dagger (\vxi \mid \delta, \lambda)$ is separable with respect to $\vxi_\ell$ and \eqref{xi ell} is strictly convex.
Hence, \cite{Tseng2001} guarantees that the solution obtained by Algorithm~\ref{alg main2} converges to the optimal solution.
Furthermore, similar to Theorem~\ref{th lam}, $\lambda_{\max}$ for $\PRSS^\dagger (\vxi \mid \delta, \lambda)$ is given by
\begin{align*}
  \lambda_{\max} (\delta)
    = \max_{\ell \in \{ 1, \ldots, k \}, j \in \{ 1, \ldots, q \}} \dfrac{|\vu' \vz_{\ell j}^\dagger|}{w_{\ell, j} (\delta)},
\end{align*}
where $\vz_{\ell j}^\dagger = (a_{1, \ell} \vh_j', \ldots, a_{n, \ell} \vh_j')'$.

%=============================
%     Sec3.3
%=============================
\subsection{Extended Hierarchical Overlapping Group Lasso}

Thus far, we have discussed hierarchical overlapping by adding parameters one by one to select the degrees of the polynomials.
Here, we extend HOGL to flexible hierarchical overlapping.
As an example, consider using the following Fourier basis instead of the polynomial basis [e.g., \citealp{vonRosen2018}, Chap.~1]:
\begin{align*}
  \vx (t)
    = \left(
         \cos (q-1) t, \sin (q-1) t, \cos (q-2) t, \sin (q-2) t, \ldots, 
           \cos t, \sin t, 1
       \right)'.
\end{align*}
For simplicity, the constants used to standardize the column vectors of $\vX$ are omitted from the expression.
Then, the varying coefficients in \eqref{beta} are given by
\begin{align*}
  \beta_\ell (t)
    = \sum_{j=1}^{q-1} \left( \cos (q - j) t, \sin (q - j) t \right) \vtheta_{\ell, j}
         + \theta_{\ell, q},
\end{align*}
where $\vtheta_{\ell, j}$ is a two-dimensional vector.
To select a pair of basis functions $(\cos (q - j) t, \sin (q - j) t)$, we need HOGL with hierarchical overlapping by adding parameters two at a time.
To address the selection of basis functions with group structure, we extend HOGL.

Here, we divide the parameter vector $\vtheta_\ell\ (\ell \in \{ 1, \ldots, k \})$ as
\begin{align*}
  \vtheta_\ell
    = \left( 
         \vtheta_{\ell, 1}', \ldots, \vtheta_{\ell, q}'
       \right)',
\end{align*}
where $\vtheta_{\ell, j}$ is an $m_j$-dimensional vector and $\vtheta_\ell$ is an $m = \sum_{j=1}^q m_j$-dimensional vector.
Furthermore, for $m$-dimensional block vector $\vgamma = (\vgamma_1', \ldots, \vgamma_q')'$, we extend the sub-vector in \eqref{subvec} to
\begin{align}
\label{subvec2}
  \vgamma_{(j)}
    = \left( \vgamma_1', \ldots, \vgamma_j' \right)',
\end{align}
where $\vgamma_j$ is an $m_j$-dimensional vector.
Since this notation naturally extends the HOGL penalty in \eqref{HOGLP}, we call the approach extended HOGL (EHOGL).
When $m_1 = \cdots = m_q = 1$, EHOGL reduces to HOGL.
To solve the optimization problem for EHOGL, the essential task is to minimize the following function:
\begin{align}
\label{f2}
  f (\vgamma)
    = \dfrac{1}{2} \| \vgamma \|^2 - \vb' \vgamma   
         + \sum_{j=1}^q \lambda_j \| \vgamma_{(j)} \|, \quad
  \vgamma
    = \left( \vgamma_1', \ldots, \vgamma_q' \right)',
\end{align}
where $\vb = (\vb_1', \ldots, \vb_q')'$ is an $m$-dimensional block vector and $\vb_j$ is an $m_j$-dimensional vector of constants.
To minimize $f (\vgamma)$, we extend $d_{\alpha, j}$ in \eqref{d} to
\begin{align}
\label{d2}
 d_{\alpha, j} &= \begin{dcases}
   0 &(j < \alpha) \\
   \left( \| \vb_j \| - \lambda_j \right)_+
     & (j = \alpha) \\
   \left( \sqrt{d_{\alpha, j-1}^2 + \| \vb_j \|^2} - \lambda_j \right)_+
     & (\alpha < j)
   \end{dcases} \quad (\alpha, j \in \{ 1, \ldots, q \}).
\end{align}
Then, the minimizer of $f (\vgamma)$ is given by the following theorem (the proof is given in Appendix~\ref{ap mainTh}).
\begin{Thm}
\label{th main2}
  Define $\alpha_\ast$ as
  \begin{align*}
    \alpha_\ast &= \begin{dcases}
        \min \cA &(\cA \neq \emptyset) \\
        q + 1 &(\cA = \emptyset)
      \end{dcases}, \quad
    \cA = \left\{
        \alpha \in \{ 1, \ldots, q \} \mid \forall j \in \{ \alpha, \alpha + 1, \ldots, q \}, d_{\alpha, j} > 0
      \right\}.
  \end{align*}
  Let $\vgamma^\ast = ({\vgamma_1^\ast}', \ldots, {\vgamma_q^\ast}')'$ be the minimizer of $f (\vgamma)$ in \eqref{f2}, i.e., $\vgamma^\ast = \arg \min_{\vgamma \in \bR^m} f (\vgamma)$.
  Then, $\vgamma_j^\ast$ is given by
  \begin{align*}
    \vgamma_j^\ast = \begin{dcases}
        \0_{m_j} &(j < \alpha_\ast) \\
        \prod_{\ell=j}^q \dfrac{d_{\alpha_\ast, \ell}}{d_{\alpha_\ast, \ell} + \lambda_\ell} \vb_j
          &(\alpha_\ast \le j)
      \end{dcases}.
  \end{align*}
\end{Thm}
Similar to Proposition~\ref{propd}, $d_{\alpha, j}$ in \eqref{d2} satisfies $d_{\alpha, j} \ge d_{\alpha + 1, j}$.
Hence, $\alpha_\ast$ in Theorem~\ref{th main2} can be obtained by Algorithm~\ref{alg d2}.

%%%%%%%%%%%%%%%%%%%%%%%%%%%%%%
%     Sec4
%%%%%%%%%%%%%%%%%%%%%%%%%%%%%%
\section{Numerical study}
\label{sec sim}

In this section, we evaluate the performance of the proposed HOGL using Monte Carlo simulation with 1000 iterations,
and compare the performance of HOGL with that of wSCAD and a third method using sparse group Lasso.
The numerical calculation programs are executed in R (ver. 4.5.0; R Core Team, 2025) on a computer running the Windows 11 Pro operating system with an AMD EPYC TM 7763 processor and 128 GB of RAM.
HOGL is available via R package \texttt{HOGLgmanova} \citep{HOGLgmanova010}.

We first describe the setting of the simulation.
The simulation model is defined by
\begin{align*}
  \vY &\sim N_{n \times p} (\vA \vTheta \vX', \vSigma \otimes \vI_n), \quad
  \vA = \vA_0 \vPsi^{1/2}, \quad
  \vPsi = \vR_k^{1/2} \vOmega_k (0.5) \vR_k^{1/2}, \quad
  \vSigma = \vR_p^{1/2} \vOmega_p (0.5) \vR_p^{1/2},
\end{align*}
where $\vA_0$ is an $n \times k$ matrix with elements identically and independently distributed according to $U (-1, 1)$, $\vR_k = \diag (1, \ldots, k)$, and $\vOmega_k (\rho)$ is a $k \times k$ autocorrelation matrix with the $(i, j)$th element $\rho^{|i-j|}$.
Furthermore, $\vX$ is a $p \times q$ matrix of polynomial basis functions of degree $(q-1)$ and the $j\ (\in \{ 1, \ldots, p \})$th time point is given by $t_j = 2 (j - 1) / (p - 1) - 1$, where $t_j$ is the $j$th of the points obtained by uniformly dividing $[-1, 1]$.
The $\vTheta$ is defined by
\begin{align*}
  \vTheta = \begin{pmatrix}
      \vO_{k_\ast, q - 6} & \vTheta_\ast \\ \vO_{k - k_\ast, q - 6} & \vO_{k - k_\ast, 6}
    \end{pmatrix}, \quad
  \vTheta_\ast = \nu \begin{pmatrix}
      0 & 0 & 0 & 0 & -3 & 0.5 \\
      0 & 0 & 0 & 4 & 1 & -2 \\
      0 & 0 & 6 & -2 & -4 & 2 \\
      0 & 12 & 3 & -12 & -3 & 1.5 \\
      -12 & -1 & 15 & 1 & -1 & -0.5      
    \end{pmatrix},
\end{align*}
where $\nu$ is a parameter adjusting the signal-to-noise ratio (SNR). The number of true explanatory variables is $k_\ast = 5$, and the true degree of varying coefficient $\beta_\ell (t)\ (\ell \in \{ 1, \ldots, k_\ast \})$ in \eqref{beta} is $\ell$.
\begin{figure}[h]
  \centering
  \includegraphics[scale=0.28]{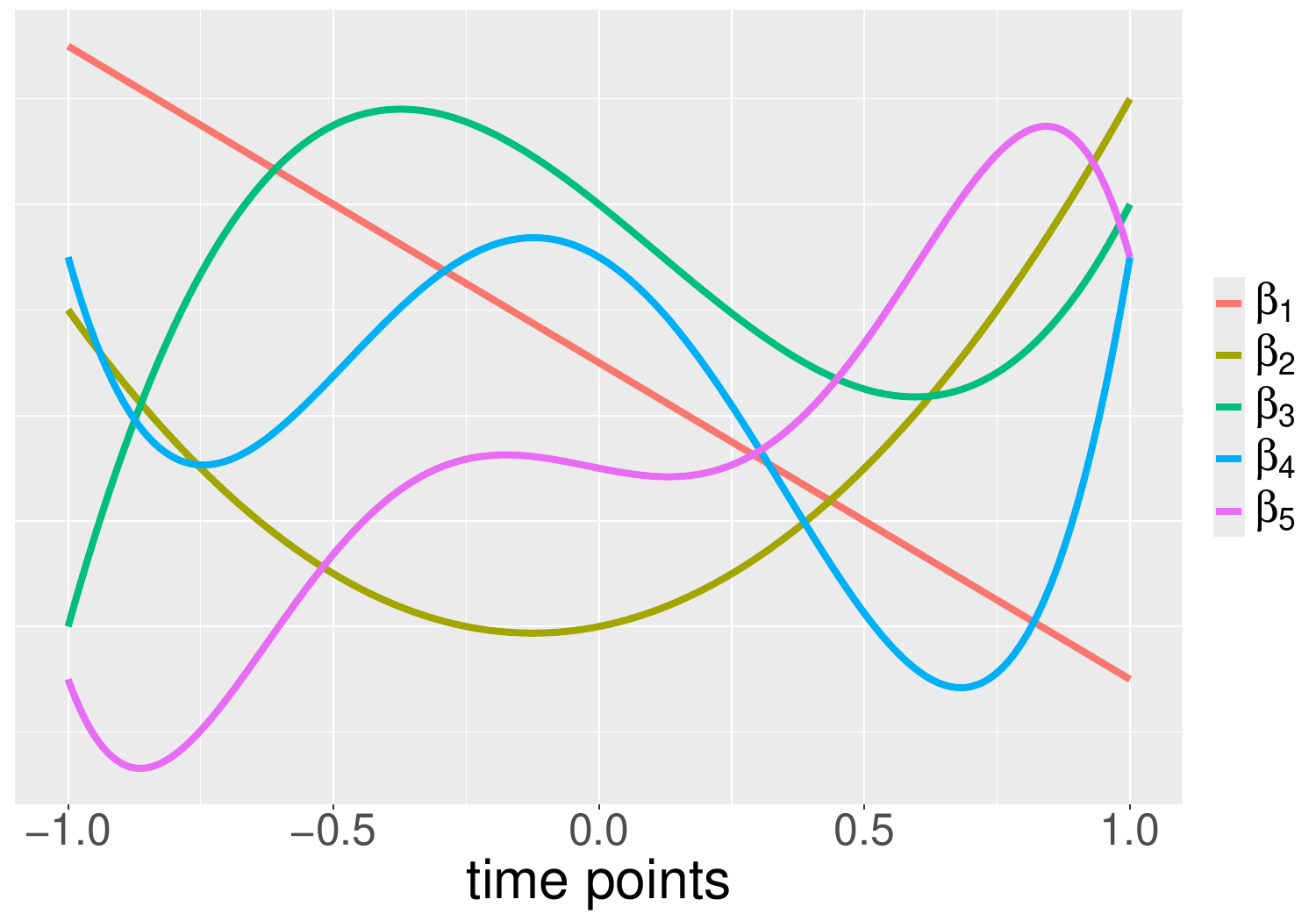}
  \caption{\ Shapes of varying coefficients $\vbeta_\ell (t)$}
  \label{fig beta}
\end{figure}
Figure~\ref{fig beta} shows the shapes of $\vbeta_\ell (t)$.
Furthermore, SNR is defined by 
\begin{align*}
  \mathrm{SNR}
    = \dfrac{1}{p} \sum_{j=1}^p \dfrac{\Var [\va' \vTheta \vx_j]}{\sigma_j^2}
    = \dfrac{1}{3p} \sum_{j=1}^p \dfrac{\vx_j' \vTheta' \vPsi \vTheta \vx_j}{\sigma_j^2},
\end{align*}
and $\nu$ is defined such that $\mathrm{SNR} = 1$, where $\va = \vPsi^{1/2} \va_0$, $\va_0$ is a $k$-dimensional vector with elements identically and independently distributed according to $U (-1, 1)$, $\vx_j$ is the $j$th row vector of $\vX$ and $\sigma_j^2$ is the $j$th diagonal element.

\begin{table}[h]
\caption{\ Selection probability (\%) when $q = 6$ \label{tab SP1}}
\centering
\begin{tabular}{rrrrrrrrrrr}
\toprule
  &  &  & \multicolumn{4}{c}{Variable} & \multicolumn{4}{c}{Degree} \\
\cmidrule(lr){4-7} \cmidrule(lr){8-11}
 $n$ & $p$ & $k$ & HOGL1 & HOGL2 & SGL & wSCAD & HOGL1 & HOGL2 & SGL & wSCAD \\
\midrule
 100 & 10 & 10 & 14.4 & \bf 79.2 & 15.4 & 2.8 & 4.5 & \bf 51.1 & 3.2 & 0.0 \\
 300 & 10 & 10 & 36.9 & \bf 93.0 & 31.2 & 23.2 & 17.5 & \bf 78.8 & 7.0 & 12.6 \\
 500 & 10 & 10 & 54.1 & \bf 97.3 & 39.6 & 81.8 & 25.7 & \bf 88.0 & 9.8 & 68.4 \\ \midrule
 100 & 40 & 10 & 2.8 & \bf 58.6 & 2.8 & 2.4 & 1.9 & \bf 37.1 & 1.0 & 0.0 \\
 300 & 120 & 10 & 35.0 & \bf 86.9 & 32.7 & 49.1 & 16.5 & \bf 73.0 & 3.3 & 36.8 \\
 500 & 200 & 10 & 61.4 & \bf 92.5 & 58.4 & 90.8 & 26.2 & \bf 80.3 & 5.2 & 79.1 \\ \midrule
 100 & 10 & 40 & 8.4 & \bf 60.3 & 2.1 & 0.0 & 0.9 & \bf 42.7 & 3.7 & 0.0 \\
 300 & 10 & 120 & 21.5 & \bf 89.8 & 9.0 & 0.0 & 3.7 & \bf 70.6 & 9.0 & 0.0 \\
 500 & 10 & 200 & 27.9 & \bf 94.3 & 18.3 & 0.4 & 5.5 & \bf 82.6 & 7.5 & 27.4 \\ \midrule
 100 & 20 & 40 & 3.7 & \bf 47.8 & 0.7 & 0.0 & 1.3 & \bf 39.3 & 1.8 & 0.0 \\
 300 & 60 & 120 & 11.7 & \bf 87.5 & 4.3 & 2.2 & 2.6 & \bf 71.3 & 6.5 & 0.3 \\
 500 & 100 & 200 & 28.8 & \bf 90.9 & 10.9 & 78.1 & 5.8 & 84.3 & 5.2 & \bf 86.9 \\
\bottomrule
\end{tabular}
\end{table}
Under the setting described above, we determine the selection probabilities of the true variables and degrees, as well as the mean squared errors (MSEs) and the runtime.
Selection probabilities are based on the percentage (\%) of the time that the true variables or degrees are selected over 1000 iterations.
MSEs are defined for the matrices of the fitted values $\hat{\vY}$ and the estimators of the regression coefficients $\hat{\vTheta}$, as shown below:
\begin{align*}
  \MSE_\mathrm{f} \left[ \hat{\vY} \right]
    &= \dfrac{1}{n p} \E \left[ 
           \tr \left\{ \left( \hat{\vY} - \vA \vTheta \vX' \right)' \left( \hat{\vY} - \vA \vTheta \vX' \right) \vSigma^{-1} \right\}
         \right], \\
  \MSE_\mathrm{c} \left[ \hat{\vTheta} \right]
    &= \dfrac{1}{kq} \E \left[ 
           \tr \left\{ \left( \hat{\vTheta} - \vTheta \right)' \left( \hat{\vTheta} - \vTheta \right) \right\}
         \right],
\end{align*}
where expectation is evaluated by Monte Carlo simulation with 1000 iterations.
The methods used in this simulation are as follows.
\begin{enumerate}[\quad$\bullet$]
\item
  HOGL1: proposed method with $\vV$ using Algorithm~\ref{alg main1}.
\item
  HOGL2: proposed method with $\vH$ using Algorithm~\ref{alg main2}, where $\vH$ is obtained by transforming $\vV$ as described in Section~\ref{sec trans}.
\item
  SGL: penalized estimation method with sparse group Lasso penalty instead of HOGL penalty in \eqref{HOGLP}.
\item
  wSCAD: method proposed by \cite{Hu-2014}.
\end{enumerate}
Based on adaptive Lasso \citep{Zou2006}, penalty weights in HOGL penalty are defined by
\begin{align*}
  w_{\ell, j}^{(0)} = \begin{dcases}
      \left\| \left( \tilde{\vtheta}_\ell \right)_{(j)} \right\|^{-1} & (\text{HOGL1}) \\
      \left\| \left( \tilde{\vxi}_\ell \right)_{(j)} \right\|^{-1} & (\text{HOGL2}) 
    \end{dcases}, \quad
  \tilde{\vtheta}_\ell
    = \tilde{\vTheta}' \ve_\ell,\ 
  \tilde{\vxi}_\ell
    = \tilde{\vXi}' \ve_\ell,
\end{align*}
where $\tilde{\vTheta}$ and $\tilde{\vXi}$ are the ordinary least squares estimators of $\vTheta$ and $\vXi$, respectively, defined by
\begin{align*}
  \tilde{\vTheta}
    = (\vA' \vA)^{-1} \vA' \vU \vV (\vV' \vV)^{-1}, \quad
  \tilde{\vXi}
    = \tilde{\vTheta} \vQ'
    = (\vA' \vA)^{-1} \vA' \vU \vH,
\end{align*}
and $\ve_\ell$ is a $k$-dimensional unit vector with the $\ell$th element one.
Regarding the two tuning parameters $\delta$ and $\lambda$, we set 10 and 100 candidates for $\delta$ and $\lambda$, respectively, and select the best pair using a grid search based on minimizing the extended GCV (EGCV) criterion \citep{Ohishi-2020} defined by
\begin{align*}
  \mathrm{EGCV} (\delta, \lambda \mid \alpha)
    = \dfrac{\tr \{ (\vY - \hat{\vY}_{\delta, \lambda})' (\vY - \hat{\vY}_{\delta, \lambda}) \vS^{-1} \}}{(1 - \df_{\delta, \lambda} / np)^\alpha},
\end{align*}
where $\hat{\vY}_{\delta, \lambda}$ is a matrix of fitted values under given $\delta$ and $\lambda$, $\df_{\delta, \lambda}$ is the number of corresponding non-zero parameters, and we set $\alpha = \log (np)$.
Similar to HOGL, we incorporate penalty weights into SGL and select two tuning parameters.
Regarding wSCAD, we define 10 candidates for each of the three tuning parameters and select the best pair by grid search based on minimizing the BIC as proposed by \cite{Hu-2014}.
Then, wSCAD is guaranteed to have the oracle properties under a large sample asymptotic framework (i.e., only $n$ diverges to infinity).
Note that $\alpha = \log (np)$ in the EGCV criterion corresponds to the BIC in \cite{Hu-2014}.
This simulation considers the following four cases as $n$ increases: (1) $p = k = 10$, (2) $p/n = 0.4$, $k = 10$, (3) $p = 10$, $k/n = 0.4$, and (4) $p/n = 0.2$, $k/n = 0.4$.

\begin{table}[h]
\caption{\ MSE when $q = 6$ \label{tab MSE1}}
\centering
\begin{tabular}{rrrrrrrrrrr}
\toprule
  &  &  & \multicolumn{4}{c}{$\MSE_\mathrm{f}$} & \multicolumn{4}{c}{$\MSE_\mathrm{c}$} \\
\cmidrule(lr){4-7} \cmidrule(lr){8-11}
 $n$ & $p$ & $k$ & HOGL1 & HOGL2 & SGL & wSCAD & HOGL1 & HOGL2 & SGL & wSCAD \\
\midrule
 100 & 10 & 10 & 0.070 & \bf 0.043 & 0.071 & 0.125 & 0.625 & \bf 0.253 & 0.548 & 1.210 \\
 300 & 10 & 10 & 0.020 & \bf 0.012 & 0.021 & 0.044 & 0.153 & \bf 0.059 & 0.176 & 0.368 \\
 500 & 10 & 10 & 0.011 & \bf 0.007 & 0.013 & 0.029 & 0.074 & \bf 0.030 & 0.108 & 0.231 \\ \midrule
 100 & 40 & 10 & 0.028 & 0.022 & 0.031 & \bf 0.022 & 1.738 & \bf 0.840 & 2.720 & 3.652 \\
 300 & 120 & 10 & 0.005 & 0.004 & 0.005 & \bf 0.003 & 0.492 & \bf 0.219 & 1.097 & 1.167 \\
 500 & 200 & 10 & 0.003 & 0.002 & 0.003 & \bf 0.001 & 0.285 & \bf 0.124 & 0.707 & 0.344 \\ \midrule
 100 & 10 & 40 & 0.291 & \bf 0.063 & 0.154 & 0.616 & 0.634 & \bf 0.099 & 0.232 & 0.523 \\
 300 & 10 & 120 & 0.102 & \bf 0.015 & 0.075 & 0.270 & 0.087 & \bf 0.007 & 0.035 & 0.042 \\
 500 & 10 & 200 & 0.063 & \bf 0.008 & 0.052 & 0.118 & 0.030 & \bf 0.002 & 0.015 & 0.017 \\ \midrule
 100 & 20 & 40 & 0.159 & \bf 0.041 & 0.089 & 0.299 & 0.995 & \bf 0.184 & 0.384 & 0.708 \\
 300 & 60 & 120 & 0.026 & \bf 0.006 & 0.020 & 0.123 & 0.227 & \bf 0.020 & 0.092 & 0.138 \\
 500 & 100 & 200 & 0.009 & \bf 0.003 & 0.008 & 0.012 & 0.062 & \bf 0.006 & 0.041 & 0.048 \\
\bottomrule
\end{tabular}
\end{table}
\begin{table}[h]
\caption{\ Selection probability (\%) when $q = 10$ \label{tab SP2}}
\centering
\begin{tabular}{rrrrrrrrrrr}
\toprule
  &  &  & \multicolumn{4}{c}{Variable} & \multicolumn{4}{c}{Degree} \\
\cmidrule(lr){4-7} \cmidrule(lr){8-11}
 $n$ & $p$ & $k$ & HOGL1 & HOGL2 & SGL & wSCAD & HOGL1 & HOGL2 & SGL & wSCAD \\
\midrule
 100 & 10 & 10 & 0.9 & \bf 74.8 & 20.7 & 0.8 & 0.0 & \bf 31.6 & 0.0 & 0.0 \\
 300 & 10 & 10 & 1.1 & \bf 91.6 & 24.7 & 0.4 & 0.0 & \bf 65.1 & 0.0 & 0.0 \\
 500 & 10 & 10 & 0.4 & \bf 95.8 & 28.5 & 0.3 & 0.0 & \bf 80.2 & 0.0 & 0.0 \\ \midrule
 100 & 40 & 10 & 0.9 & \bf 53.2 & 16.5 & 1.0 & 0.0 & \bf 19.7 & 0.0 & 0.0 \\
 300 & 120 & 10 & 0.1 & \bf 87.0 & 42.1 & 1.0 & 0.0 & \bf 61.1 & 0.0 & 0.0 \\
 500 & 200 & 10 & 0.0 & \bf 92.4 & 50.6 & 1.3 & 0.0 & \bf 76.1 & 0.0 & 0.1 \\ \midrule
 100 & 10 & 40 & 0.0 & \bf 52.4 & 5.1 & 0.0 & 0.0 & \bf 20.6 & 0.0 & 0.0 \\
 300 & 10 & 120 & 0.2 & \bf 85.7 & 13.3 & 0.0 & 0.0 & \bf 49.6 & 0.0 & 0.0 \\
 500 & 10 & 200 & 0.1 & \bf 89.8 & 11.8 & 0.0 & 0.0 & \bf 67.2 & 0.0 & 0.0 \\ \midrule
 100 & 20 & 40 & 0.1 & \bf 41.3 & 4.6 & 0.0 & 0.0 & \bf 20.6 & 0.0 & 0.0 \\
 300 & 60 & 120 & 0.0 & \bf 83.5 & 9.7 & 0.0 & 0.0 & \bf 53.7 & 0.0 & 0.0 \\
 500 & 100 & 200 & 0.1 & \bf 89.4 & 13.0 & 4.5 & 0.0 & \bf 74.3 & 0.0 & 0.0 \\
\bottomrule
\end{tabular}
\end{table}
Table~\ref{tab SP1} summarizes the selection probabilities of the true variables and degrees when $q=6$, i.e., fitting a polynomial of degree five.
We can see that HOGL2 performed very well in both variable selection and degree selection.
Although wSCAD also performed well, the performance declined when $k$ increased.
It can be considered that the theoretical properties of wSCAD appear in the results.
The performances of HOGL1 and SGL were not good; in particular, degree selection was quite poor.

Table~\ref{tab MSE1} summarizes the MSEs of the fitted values and coefficients when $q=6$.
Regarding the MSE of the coefficients, HOGL2 was always superior.
We can consider that this was caused by the high performances of variable selection and degree selection.
HOGL2 also performed well with respect to the MSE of the fitted values.
However, in the case in which $n$ and $p$ increase and $k$ is fixed, wSCAD performed best, although differences among the four methods became small.
It can be guessed that this setting represents a desirable situation in which sample size and number of time points are sufficiently large for the number of explanatory variables and the dimension of the polynomial basis.
In addition, since the tuning parameters are selected according to prediction accuracy, the MSE of the fitted values may be good even when the performances of variable selection and degree selection are poor.

Table~\ref{tab SP2} summarizes the probabilities of selecting the true variables and degrees when $q=10$, i.e., fitting a polynomial of degree nine.
Comparing results with the case in which $q=6$, HOGL2 maintained its high performance; however, the performances of the other methods declined.
It can be considered that fitting a high-degree polynomial rendered the model unstable, which made identification of the basis vectors difficult.
\begin{figure}[h]
  \centering
  \includegraphics[scale=0.28]{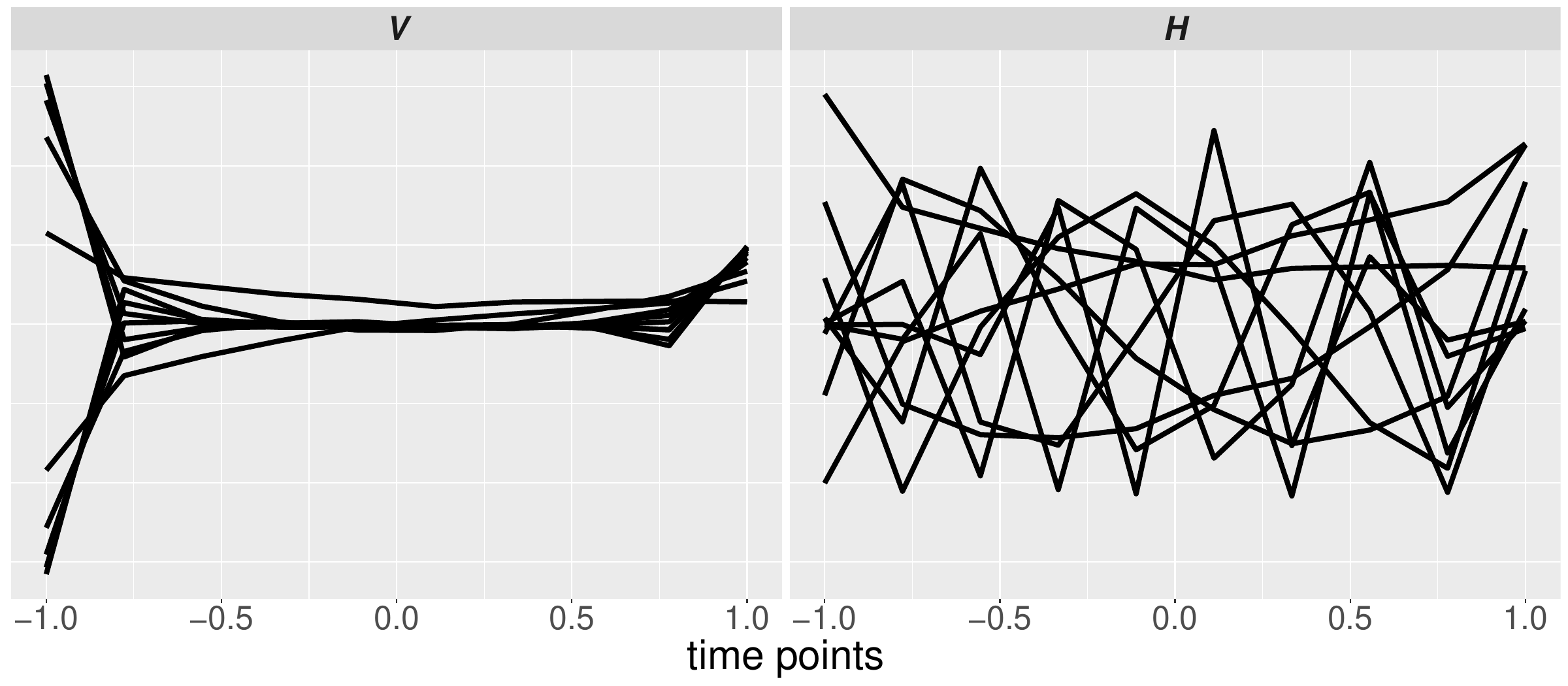}
  \caption{\ Standardized polynomial basis functions (left) and transformed basis functions (right)}
  \label{fig VH}
\end{figure}
Figure~\ref{fig VH} shows the basis functions when $n=500$, $p=10$, $k=10$, and $q=10$; the left and right panels show the column vectors of $\vV$ and $\vH$, respectively.
From the figure, we can see that although identification of the basis functions of the original polynomial basis is difficult, basis function differences became very clear when transforming the polynomial basis to an orthonormal basis.
Hence, we can conclude that the performances of HOGL1, SGL, and wSCAD, which use $\vV$, declined, while HOGL2, which uses $\vH$, was not affected by high-degree polynomials.

\begin{table}[h]
\caption{\ MSE when $q = 10$ \label{tab MSE2}}
\centering
\begin{tabular}{rrrrrrrrrrr}
\toprule
  &  &  & \multicolumn{4}{c}{$\MSE_\mathrm{f}$} & \multicolumn{4}{c}{$\MSE_\mathrm{c}$} \\
\cmidrule(lr){4-7} \cmidrule(lr){8-11}
 $n$ & $p$ & $k$ & HOGL1 & HOGL2 & SGL & wSCAD & HOGL1 & HOGL2 & SGL & wSCAD \\
\midrule
 100 & 10 & 10 & 0.334 & \bf 0.049 & 0.216 & 0.670 & 2616.288 & \bf 0.857 & 3518.354 & 884.144 \\
 300 & 10 & 10 & 0.043 & \bf 0.013 & 0.053 & 0.071 & 1245.160 & \bf 0.056 & 1142.321 & 1267.022 \\
 500 & 10 & 10 & 0.026 & \bf 0.007 & 0.030 & 0.036 & 733.753 & \bf 0.027 & 672.788 & 753.098 \\ \midrule
 100 & 40 & 10 & 0.050 & \bf 0.024 & 0.049 & 0.076 & 1696.813 & \bf 2.101 & 1569.357 & 1271.156 \\
 300 & 120 & 10 & 0.008 & \bf 0.004 & 0.007 & 0.005 & 858.507 & \bf 0.425 & 743.454 & 646.727 \\
 500 & 200 & 10 & 0.004 & 0.002 & 0.003 & \bf 0.002 & 576.993 & \bf 0.134 & 489.744 & 431.723 \\ \midrule
 100 & 10 & 40 & 0.987 & \bf 0.074 & 0.511 & 0.790 & 34.732 & \bf 0.106 & 737.015 & 0.703 \\
 300 & 10 & 120 & 0.734 & \bf 0.018 & 0.366 & 0.774 & 1.505 & \bf 0.008 & 106.320 & 2.961 \\
 500 & 10 & 200 & 0.627 & \bf 0.009 & 0.306 & 0.743 & 0.734 & \bf 0.002 & 41.426 & 2.731 \\ \midrule
 100 & 20 & 40 & 0.397 & \bf 0.047 & 0.274 & 0.523 & 5.060 & \bf 0.164 & 163.205 & 12.746 \\
 300 & 60 & 120 & 0.126 & \bf 0.006 & 0.061 & 0.365 & 2.409 & \bf 0.021 & 58.196 & 10.226 \\
 500 & 100 & 200 & 0.076 & \bf 0.003 & 0.033 & 0.242 & 1.322 & \bf 0.005 & 27.600 & 10.187 \\
\bottomrule
\end{tabular}
\end{table}
Table~\ref{tab MSE2} summarizes the MSEs for the fitted values and coefficients when $q=10$.
As can be seen here, the MSE values for the coefficients clearly reflect the performances of variable and degree selections, while the MSEs for the fitted values are barely affected.
\begin{table}[h]
\caption{\ Example of estimation results \label{tab example}}
\centering
\begin{tabular}{crrrrrrrrrrrr}
\toprule
  & \multicolumn{2}{c}{Estimation error} & \multicolumn{10}{c}{Estimated degrees} \\
\cmidrule(lr){2-3} \cmidrule(lr){4-13}
 & Fitted values & Coefficients & $\beta_{1}$ & $\beta_{2}$ & $\beta_{3}$ & $\beta_{4}$ & $\beta_{5}$ & $\beta_{6}$ & $\beta_{7}$ & $\beta_{8}$ & $\beta_{9}$ & $\beta_{10}$ \\
\midrule
 HOGL1 & 0.003 & 580.863 & 9 & 9 & 9 & 9 & 9 & 0 & 9 & 9 & 9 & 9 \\
 HOGL2 & 0.002 & 0.050 & 1 & 2 & 3 & 4 & 5 & 0 & 0 & 0 & 0 & 0 \\
 SGL & 0.003 & 398.871 & 9 & 9 & 9 & 9 & 9 & 0 & 0 & 0 & 0 & 0 \\
 wSCAD & 0.001 & 359.984 & 1 & 9 & 9 & 9 & 9 & 0 & 8 & 9 & 9 & 0 \\
\bottomrule
\end{tabular}
\end{table}
\begin{figure}[h]
  \centering
  \includegraphics[scale=0.28]{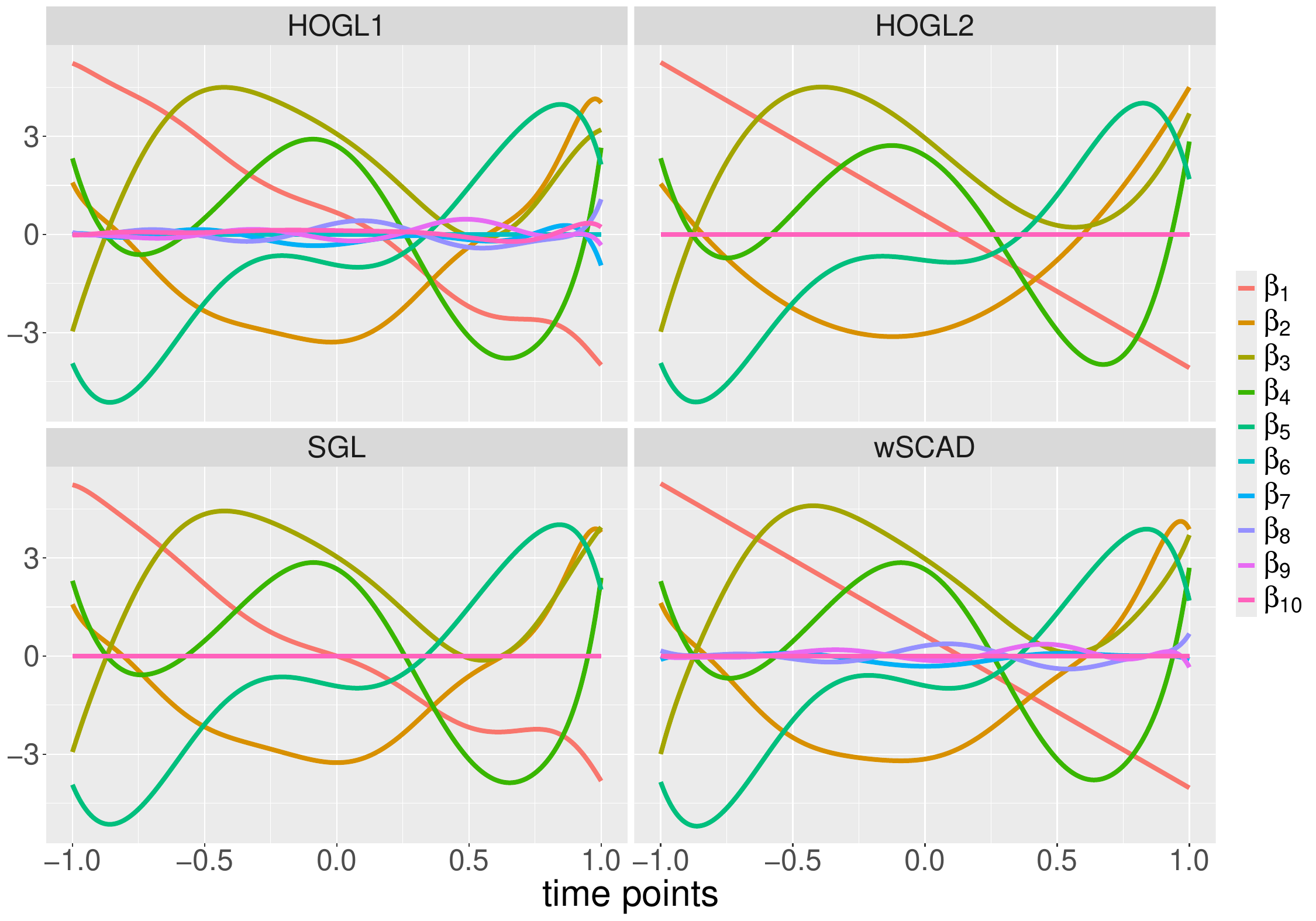}
  \caption{\ Example of estimated varying coefficients}
  \label{fig example}
\end{figure}
Table~\ref{tab example} and Figure~\ref{fig example} show one example of estimation results when $n=500$, $p=200$, $k=10$, and $q=10$, where the estimation errors of  the matrices of fitted values $\hat{\vY}$ and the estimates of regression coefficients $\hat{\vTheta}$ are given by $\tr \{ (\hat{\vY} - \vA \vTheta \vX')' (\hat{\vY} - \vA \vTheta \vX') \vSigma^{-1} \} / np$ and $\tr \{ (\hat{\vTheta} - \vTheta)' (\hat{\vTheta} - \vTheta) \} / kq$, respectively, and $\vbeta_\ell$ is given by \eqref{beta}.
In this example, only HOGL2 was able to perfectly select the true variables and degrees, and the estimation error for the coefficients was very small.
On the other hand, since the other methods were unable to make similarly accurate selections, the estimation errors of the coefficients were much larger.
However, the differences among the estimation errors for the fitted values were relatively small, as can be seen in Figure~\ref{fig example}.
The figure shows that the shapes of the varying coefficients do not exhibit large differences even when the true degrees were not selected, which would explain why the incorrect selection of the true degrees did not materially affect the estimation error of the fitted values.
Nevertheless, correct selection of the true degrees is important to properly and concisely interpret a model.

\begin{table}[h]
\caption{\ Runtime (sec.) \label{tab time}}
\centering
\begin{tabular}{rrrrrrrrrrr}
\toprule
  &  &  & \multicolumn{4}{c}{$q=6$} & \multicolumn{4}{c}{$q=10$} \\
\cmidrule(lr){4-7} \cmidrule(lr){8-11}
 $n$ & $p$ & $k$ & HOGL1 & HOGL2 & SGL & wSCAD & HOGL1 & HOGL2 & SGL & wSCAD \\
\midrule
 100 & 10 & 10 & 135.6 & \bf 5.0 & 10.5 & 15.8 & 127.0 & 7.5 & \bf 6.8 & 18.3 \\
 300 & 10 & 10 & 127.5 & \bf 4.6 & 10.1 & 15.1 & 140.6 & \bf 6.8 & 7.4 & 18.1 \\
 500 & 10 & 10 & 116.7 & \bf 4.4 & 10.5 & 14.6 & 150.4 & \bf 6.6 & 7.9 & 18.5 \\ \midrule
 100 & 40 & 10 & 148.2 & \bf 5.3 & 9.5 & 14.4 & 165.8 & 7.3 & \bf 6.3 & 18.6 \\
 300 & 120 & 10 & 119.7 & \bf 5.0 & 16.7 & 14.5 & 171.5 & \bf 6.8 & 24.1 & 19.6 \\
 500 & 200 & 10 & 95.4 & \bf 6.3 & 43.1 & 17.1 & 183.9 & \bf 7.6 & 66.9 & 20.5 \\ \midrule
 100 & 10 & 40 & 1272.5 & \bf 35.5 & 137.6 & 67.9 & 1638.8 & \bf 56.2 & 76.6 & 87.6 \\
 300 & 10 & 120 & 5485.4 & \bf 102.8 & 909.2 & 405.1 & 15193.4 & \bf 165.7 & 662.8 & 1064.8 \\
 500 & 10 & 200 & 8617.4 & \bf 166.6 & 2656.5 & 1339.6 & 44570.6 & \bf 273.2 & 2064.3 & 4726.5 \\ \midrule
 100 & 20 & 40 & 1325.8 & \bf 36.3 & 97.0 & 60.4 & 1996.5 & 53.8 & \bf 40.5 & 79.3 \\
 300 & 60 & 120 & 5413.2 & \bf 85.0 & 946.5 & 361.5 & 17647.3 & \bf 130.9 & 530.2 & 917.0 \\
 500 & 100 & 200 & 5899.0 & \bf 121.4 & 3610.8 & 1212.9 & 49388.0 & \bf 180.8 & 2715.4 & 4149.9 \\
\bottomrule
\end{tabular}
\end{table}
Table~\ref{tab time} summarizes runtime.
The table shows that HOGL2 has advantages not only in terms of model selection and MSE but also in calculation speed.
In contrast, HOGL1 is particularly slow.
Thus, we can say that transformation of the matrix of basis functions affects not only the performance of model selection but also the calculation speed.

%%%%%%%%%%%%%%%%%%%%%%%%%%%%%%
%     Sec5
%%%%%%%%%%%%%%%%%%%%%%%%%%%%%%
\section{Concluding remarks}

This paper proposed hierarchical overlapping group Lasso (HOGL) to perform the selections of the explanatory variables and the degrees of the polynomial basis functions simultaneously in the GMANOVA model.
By hierarchically applying group Lasso to coefficients, beginning with higher-order coefficients, HOGL can uniformly address both selection problems.
Algorithms with optimality and convergence that can be used to solve the optimization problem for HOGL were also proposed, based on the MM philosophy and the block-wise coordinate descent method in which the update equation of a solution is given in closed form.
In a numerical simulation comparing the proposed method with several existing methods, HOGL was shown to have advantages in terms of both variable section and degree selection, as well as MSE and calculation speed.
Notably, HOGL was able to maintain high performance even when fitting high-degree polynomials by transforming the matrix of the polynomial basis functions.
It should be noted, however, that all the methods examined have the potential for improvement.

While wSCAD is guaranteed to have oracle properties under specific conditions, appropriate candidates and search ranges for its three tuning parameters are not provided.
In the simulation, we selected the three tuning parameters from 1000 sets of candidates.
Although we would expect improvements in performance if more candidates are included, runtime requirements may render this impractical.
The two tuning parameters of HOGL were selected from 1000 pairs of candidates.
Although HOGL showed good numerical performance by selecting the optimal pair of tuning parameters using the EGCV criterion with $\alpha=\log (np)$, theoretical reasons were not offered.
The two tuning parameters of SGL were also selected from 1000 pairs of candidates  using the EGCV criterion with $\alpha = \log (np)$. However, the choice of $\alpha$ produced a large difference in performance, particularly for SGL.
The choice of $\alpha$ involved a clear trade-off between model selection performance and MSE. 
For example, $\alpha = \sqrt{np}$ improved model selection but worsened MSE, while $\alpha = \log (n p)$ was good for MSE.

Moreover, the true model in the simulation depended on SNR.
SNR can be interpreted as the difficulty of model selection: a larger SNR value implies that model selection is easier.
In fact, we found that the performances of all the methods improved and that the differences among the methods became small by setting $\mathrm{SNR} = 3$ (the results are given in Appendix~\ref{ap SNR3}).
At the same time, it is a great advantage that HOGL performed well in the setting where $\mathrm{SNR} = 1$, which corresponds to the more difficult setting.
As future work, we plan to investigate theoretical characteristics such as oracle properties and develop methods using other penalty versions (e.g., SCAD) of HOGL.

%%%%%%%%%%%%%%%%%%%%%%%%%%%%%%
%     Acknowledgement
%%%%%%%%%%%%%%%%%%%%%%%%%%%%%%
\paragraph{Acknowledgements}

The authors thank FORTE Science Communications (\url{https://www.forte-science.co.jp/}) for English language editing.
This work was partially supported by JSPS KAKENHI Grant Numbers 23H00809, 25K17296, and 25K21159.

%%%%%%%%%%%%%%%%%%%%%%%%%%%%%%
%     References
%%%%%%%%%%%%%%%%%%%%%%%%%%%%%%

\bibliographystyle{unsrtnat}
\bibliography{bibliography}  %%% Uncomment this line and comment out the ``thebibliography'' section below to use the external .bib file (using bibtex) .

\begin{thebibliography}{24}
\providecommand{\natexlab}[1]{#1}
\providecommand{\url}[1]{\texttt{#1}}
\expandafter\ifx\csname urlstyle\endcsname\relax
  \providecommand{\doi}[1]{doi: #1}\else
  \providecommand{\doi}{doi: \begingroup \urlstyle{rm}\Url}\fi

\bibitem[Potthoff and Roy(1964)]{PotthoffRoy1964}
R.~F. Potthoff and S.~N. Roy.
\newblock A generalized multivariate analysis of variance model useful
  especially for growth curve problems.
\newblock \emph{Biometrika}, 51:\penalty0 313--325, 1964.

\bibitem[Srivastava(2002)]{Srivastava2002}
M.~S. Srivastava.
\newblock \emph{Methods of Multivariate Statistics}.
\newblock John Wiley \& Sons, Inc, New York, 2002.

\bibitem[Timm(2002)]{Timm2002}
N.~H. Timm.
\newblock \emph{Applied Multivariate Analysis}.
\newblock Springer-Verlag, New York, 2002.

\bibitem[von Rosen(1991)]{vonRosen1991}
D.~von Rosen.
\newblock The growth curve model: A review.
\newblock \emph{Comm. Statist. Theory Methods}, 20:\penalty0 2791--2822, 1991.

\bibitem[Kshirsagar and Smith(1995)]{KshirsagarSmith1995}
A.~M. Kshirsagar and W.~B. Smith.
\newblock \emph{Growth Curves}.
\newblock Marcel Dekker, Inc, New York, Basel, Hong Kong, 1995.

\bibitem[Satoh and Yanagihara(2010)]{SatohYanagihara2010}
K.~Satoh and H.~Yanagihara.
\newblock Estimation of varying coefficients for a growth curve model.
\newblock \emph{Amer. J. Math. Management Sci.}, 30:\penalty0 243--256, 2010.

\bibitem[Fujikoshi and Rao(1991)]{FujikoshiRao1991}
Y.~Fujikoshi and C.~R. Rao.
\newblock Selection of covariables in the growth curve model.
\newblock \emph{Biometrika}, 78:\penalty0 779--785, 1991.

\bibitem[Satoh et~al.(1997)Satoh, Kobayashi, and Fujikoshi]{Satoh-1997}
K.~Satoh, M.~Kobayashi, and Y.~Fujikoshi.
\newblock Variable selection for the growth curve model.
\newblock \emph{J. Multivariate Anal.}, 60:\penalty0 277--292, 1997.

\bibitem[Enomoto et~al.(2015)Enomoto, Sakurai, and Fujikoshi]{Enomoto-2015}
R.~Enomoto, T.~Sakurai, and Y.~Fujikoshi.
\newblock Consistency properties of {AIC}, {BIC}, {$C_p$} and their
  modifications in the growth curve model under a large-$(q, n)$ framework.
\newblock \emph{SUT J. Math.}, 51:\penalty0 59--81, 2015.

\bibitem[Tibshirani(1996)]{Tibshirani1996}
R.~Tibshirani.
\newblock Regression shrinkage and selection via the {L}asso.
\newblock \emph{J. Roy. Statist. Soc. Ser. B}, 58:\penalty0 267--288, 1996.
\newblock \doi{10.1111/j.2517-6161.1996.tb02080.x}.

\bibitem[Tibshirani et~al.(2005)Tibshirani, Saunders, Rosset, Zhu, and
  Knight]{Tibshirani-2005}
R.~Tibshirani, M.~Saunders, S.~Rosset, J.~Zhu, and K.~Knight.
\newblock Sparsity and smoothness via the fused {L}asso.
\newblock \emph{J. R. Stat. Soc. Ser. B Stat. Methodol.}, 67:\penalty0 91--108,
  2005.
\newblock \doi{10.1111/j.1467-9868.2005.00490.x}.

\bibitem[Yuan and Lin(2006)]{YuanLin2006}
M.~Yuan and Y.~Lin.
\newblock Model selection and estimation in regression with grouped variables.
\newblock \emph{J. R. Stat. Soc. Ser. B Stat. Methodol.}, 68:\penalty0 49--67,
  2006.
\newblock \doi{10.1111/j.1467-9868.2005.00532.x}.

\bibitem[Simon et~al.(2013)Simon, Friedman, Hastie, and Tibshirani]{Simon-2013}
N.~Simon, J.~Friedman, T.~Hastie, and R.~Tibshirani.
\newblock A sparse-group {L}asso.
\newblock \emph{J. Comput. Graph. Statist.}, 22:\penalty0 231--245, 2013.
\newblock \doi{10.1080/10618600.2012.681250}.

\bibitem[Obozinski et~al.(2008)Obozinski, Wainwright, and
  Jordan]{Obozinski-2008}
G.~R. Obozinski, M.~J. Wainwright, and M.~Jordan.
\newblock High-dimensional support union recovery in multivariate regression.
\newblock In D.~Koller, D.~Schuurmans, Y.~Bengio, and L.~Bottou, editors,
  \emph{Advances in Neural Information Processing Systems}, volume~21. Curran
  Associates, Inc., 2008.

\bibitem[Yanagihara and Oda(2021)]{YanagiharaOda2021}
H.~Yanagihara and R.~Oda.
\newblock Coordinate descent algorithm for normal-likelihood-based group
  {L}asso in multivariate linear regression.
\newblock In I.~Czarnowski, R.~J. Howlett, and L.~C. Jain, editors,
  \emph{Intelligent Decision Technologies}, pages 429--439, Singapore, 2021.
  Springer Singapore.
\newblock \doi{10.1007/978-981-16-2765-1\_36}.

\bibitem[Hu et~al.(2014)Hu, Xin, and You]{Hu-2014}
J.~Hu, X.~Xin, and J.~You.
\newblock Model determination and estimation for the growth curve model via
  group {SCAD} penalty.
\newblock \emph{J. Multivariate Anal.}, 124:\penalty0 199--213, 2014.

\bibitem[Fan and Li(2001)]{FanLi2001}
J.~Fan and R.~Li.
\newblock Variable selection via nonconcave penalized likelihood and its oracle
  properties.
\newblock \emph{J. Amer. Statist. Assoc.}, 96:\penalty0 1348--1360, 2001.
\newblock \doi{10.1198/016214501753382273}.

\bibitem[Hunter and Lange(2004)]{HunterLange2004}
D.~R. Hunter and K.~Lange.
\newblock A tutorial on {MM} algorithms.
\newblock \emph{Amer. Statist.}, 58:\penalty0 30--37, 2004.
\newblock \doi{10.1198/0003130042836}.

\bibitem[Razaviyayn et~al.(2013)Razaviyayn, Hong, and Luo]{Razaviyayn-2013}
M.~Razaviyayn, M.~Hong, and Z.-Q. Luo.
\newblock A unified convergence analysis of block successive minimization
  methods for nonsmooth optimization.
\newblock \emph{SIAM J. Optim.}, 23:\penalty0 1126--1153, 2013.

\bibitem[Tseng(2001)]{Tseng2001}
P.~Tseng.
\newblock Convergence of a block coordinate descent method for
  nondifferentiable minimization.
\newblock \emph{J. Optim. Theory Appl.}, 109:\penalty0 475--494, 2001.

\bibitem[von Rosen(2018)]{vonRosen2018}
D.~von Rosen.
\newblock \emph{Bilinear Regression Analysis: An introduction}.
\newblock Springer, Cham, 2018.

\bibitem[Ohishi(2025)]{HOGLgmanova010}
M.~Ohishi.
\newblock \emph{HOGLgmanova: Variable and basis function selections in GMANOVA
  model}, 2025.
\newblock URL \url{https://github.com/ohishim/HOGLgmanova}.
\newblock {R} package version 0.1.0.

\bibitem[Zou(2006)]{Zou2006}
H.~Zou.
\newblock The adaptive {L}asso and its oracle properties.
\newblock \emph{J. Amer. Statist. Assoc.}, 101:\penalty0 1418--1429, 2006.
\newblock \doi{10.1198/016214506000000735}.

\bibitem[Ohishi et~al.(2020)Ohishi, Yanagihara, and Fujikoshi]{Ohishi-2020}
M.~Ohishi, H.~Yanagihara, and Y.~Fujikoshi.
\newblock A fast algorithm for optimizing ridge parameters in a generalized
  ridge regression by minimizing a model selection criterion.
\newblock \emph{J. Statist. Plann. Inference}, 204:\penalty0 187--205, 2020.
\newblock \doi{10.1016/j.jspi.2019.04.010}.

\end{thebibliography}





% \end{thebibliography}

%%%%%%%%%%%%%%%%%%%%%%%%%%%%%%
%     Appendix 
%%%%%%%%%%%%%%%%%%%%%%%%%%%%%%
\appendix
\section*{Appendix}
\setcounter{section}{1}
\setcounter{equation}{0}
\setcounter{table}{0}
\renewcommand{\theequation}{{\rm A}.\arabic{equation}}
\numberwithin{Lem}{section}
\numberwithin{table}{section}

%=============================
% Appendix A.1
%=============================
\subsection{Proof of Proposition~\ref{propd}}
\label{ap propd}

Notice that it is obvious when $j \le q - 2 \wedge j+1 \le \alpha \le q - 1$ and $1 \le j = \alpha \le q-1$.
Hence, we prove the following two cases:
\begin{align*}
  d_{\alpha, j} - d_{\alpha+1, j} = \begin{dcases}
      \left( \sqrt{d_{\alpha, \alpha}^2 + b_{\alpha+1}^2} - \lambda_{\alpha+1} \right)_+
        - \left( |b_{\alpha+1}| - \lambda_{\alpha+1} \right)_+ 
          & (\text{C1: } 2 \le \alpha + 1 = j \le q) \\
      \left( \sqrt{d_{\alpha, j-1}^2 + b_j^2} - \lambda_j \right)_+
        - \left( \sqrt{d_{\alpha+1, j-1}^2 + b_j^2} - \lambda_j \right)_+ 
          & (\text{C2: } 3 \le j \wedge \alpha \le j - 2) \\
    \end{dcases}.
\end{align*}
When C1, since $(d_{\alpha, \alpha}^2 + b_{\alpha+1}^2)^{1/2} \ge |b_{\alpha+1}|$, we have $d_{\alpha, \alpha+1} \ge d_{\alpha+1, \alpha+1}$.
Regarding C2, given the result for C1, it follows for $j = \alpha + 2$ that
\begin{align*}
  d_{\alpha, \alpha+2}
    = \left( \sqrt{d_{\alpha, \alpha + 1}^2 + b_{\alpha+2}^2} - \lambda_{\alpha+2} \right)_+
    \ge \left( \sqrt{d_{\alpha + 1, \alpha + 1}^2 + b_{\alpha+2}^2} - \lambda_{\alpha+2} \right)_+
    = d_{\alpha+1, \alpha+2}.
\end{align*}
Suppose that $d_{\alpha, j_0} \ge d_{\alpha+1, j_0}$ holds for $j = j_0 \in \{ \alpha+2, \alpha+3, \ldots, q-1 \}$.
Then, we have
\begin{align*}
  d_{\alpha, j_0+1} 
    = \left( \sqrt{d_{\alpha, j_0}^2 + b_{j_0+1}^2} - \lambda_{j_0+1} \right)_+
    \ge \left( \sqrt{d_{\alpha + 1, j_0}^2 + b_{j_0+1}^2} - \lambda_{j_0+1} \right)_+
    = d_{\alpha+1, j_0+1}.
\end{align*}
Hence, mathematical induction leads to $d_{\alpha, j} \ge d_{\alpha+1, j}$ for C2; consequently, Proposition~\ref{propd} is proved.

%=============================
% Appendix A.2
%=============================
\subsection{Proofs of Theorems~\ref{th main} and \ref{th main2}}
\label{ap mainTh}

We prove Theorem~\ref{th main2}.
The proof of Theorem~\ref{th main} follows by setting $m_1 = \cdots = m_q = 1$.
We define a set of block vectors as
\begin{align*}
  \bR^{m_1 + \cdots + m_q}
    = \left\{ \vx = \left( \vx_1', \ldots, \vx_q' \right)' \in \bR^m \mid \vx_j \in \bR^{m_j}\ (j = 1, \ldots, q)  \right\}.
\end{align*}
Note that although $\bR^m = \bR^{m_1 + \cdots + m_q}$ holds, we use the notation $\bR^{m_1 + \cdots + m_q}$ to emphasize the sizes of each block.
For a block vector $\vx \in \bR^{m_1 + \cdots + m_q}$, we define sub-vector $\vx_{[a:b]} \in \bR^{m_a + \cdots + m_b}$ as
\begin{align*}
  \vx_{[a:b]}
    = \left( \vx_{a}', \vx_{a+1}', \ldots, \vx_b' \right)' \quad
  (1 \le a \le b \le q).
\end{align*}
Note that $\vx_{(j)} = \vx_{[1, j]}$ holds, where $\vx_{(j)}$ is a sub-vector of $\vx$ given by \eqref{subvec2}.
The following lemma is a key to proving the theorem (the proof is given in Appendix~\ref{ap lem1}).
\begin{Lem}
\label{lem1}
  Consider the following system of equations for $\vgamma \in \bR^{m_1 + \cdots + m_q}$.
  \begin{align}
  \label{s-eq}
    \left( 1 + \sum_{\ell=j}^q \dfrac{\lambda_\ell}{\| \vgamma_{(\ell)} \|} \right) \vgamma_j = \vb_j
      \quad (\vgamma_1 \neq \0_{m_1};\ j = 1, \ldots, q).
  \end{align}
  A necessary and sufficient condition such that the system of equations has a real root is given by
  \begin{align*}
    \forall j \in \{ 1, \ldots, q \}, d_j > 0,
  \end{align*}
  and the real root is given by
  \begin{align*}
    \vgamma_j
      = \prod_{\ell=j}^q \dfrac{d_\ell}{d_\ell + \lambda_\ell} \vb_j, \quad
    d_j = \begin{dcases}
        \| \vb_1 \| - \lambda_1 &(j = 1) \\
        \sqrt{d_{j-1}^2 + \| \vb_j \|^2} - \lambda_j &(j = 2, \ldots, q)
      \end{dcases}.
  \end{align*}
\end{Lem}

We redefine $f$ in \eqref{f2} as
\begin{align*}
  f (\vgamma \mid \vb, \vlambda)
    = \dfrac{1}{2} \| \vgamma \|^2 - \vb' \vgamma
         + \sum_{j=1}^q \lambda_j \| \vgamma_{(j)} \|, \quad
  \vlambda = (\lambda_1, \ldots, \lambda_q)',
\end{align*}
and define two sets of block vectors as 
\begin{align*}
  \cB_q^{m_1 + \cdots + m_q} &= \left\{ \vx \in \bR^{m_1 + \cdots + m_q} \mid \vx_1 \neq \0_{m_1} \right\}, \quad
  \cR_{q, \alpha}^{m_1 + \cdots + m_q} = \left\{ \vx \in \bR^{m_1 + \cdots + m_q} \mid \vx_{[1:\alpha]} = \0_{m_1 + \cdots + m_\alpha} \right\}.
\end{align*}
For all $\vgamma \in \cR_{q, \alpha}^{m_1 + \cdots + m_q}$, we have
\begin{align*}
  f (\vgamma \mid \vb, \vlambda)
    &= \dfrac{1}{2} \| \vgamma_{[\alpha+1:q]} \|^2 - \vb_{[\alpha+1:q]}' \vgamma_{[\alpha+1:q]}
         + \sum_{j=1}^{q-\alpha} \lambda_{\alpha+j} \| (\vgamma_{[\alpha+1:q]})_{(j)} \| \\
    &= f (\vgamma_{[\alpha+1:q]} \mid \vb_{[\alpha+1:q]}, \vlambda_{[\alpha+1:q]}).
\end{align*}
Furthermore, $f$ is differentiable for $\vgamma \in \cB_q^{m_1 + \cdots + m_q}$ and its partial derivative is given by
\begin{align*}
  \nabla_j f (\vgamma \mid \vb, \vlambda)
    = \dfrac{\partial f}{\partial \vgamma_j} (\vgamma \mid \vb, \vlambda)
    = \vgamma_j - \vb_j + \sum_{\ell=j}^q \dfrac{\lambda_\ell}{\| \vgamma_{(\ell)} \|} \vgamma_j
    = \left( 1 + \sum_{\ell=j}^q \dfrac{\lambda_\ell}{\| \vgamma_{(\ell)} \|}  \right) \vgamma_j - \vb_j.
\end{align*}

For all $\alpha \in \{ 1, \ldots, q \}$, $f (\vgamma_{[\alpha:q]} \mid \vb_{[\alpha:q]}, \vlambda_{[\alpha:q]})$ is differentiable for $\vgamma_{[\alpha:q]} \in \cB_{q-\alpha+1}^{m_\alpha + \cdots m_q}$.
Let $\vs_\alpha$ be a root of the following system of equations when $\vgamma_{[\alpha:q]} \in \cB_{q-\alpha+1}^{m_\alpha + \cdots + m_q}$:
\begin{align*}
  \nabla_j f (\vgamma_{[\alpha:q]} \mid \vb_{[\alpha:q]}, \vlambda_{[\alpha:q]}) = \0_{m_j}
    \quad (j = \alpha, \alpha + 1, \ldots, q).
\end{align*}
Lemma~\ref{lem1} tells us whether $\vs_\alpha$ exists and $\vs_\alpha$ is given in closed form if it exists.
Notice that $f$ is a strictly convex function.
Hence, we have
\begin{align*}
\begin{dcases}
  \text{$\vs_\alpha \in \cB_{q-\alpha+1}^{m_\alpha + \cdots + m_q}$ exists}
    &\Longrightarrow \vs_\alpha = \arg \min_{\vgamma_{[\alpha:q]}} f (\vgamma_{[\alpha:q]} \mid \vb_{[\alpha:q]}, \vlambda_{[\alpha:q]}) \\
  \text{$\vs_\alpha \in \cB_{q-\alpha+1}^{m_\alpha + \cdots + m_q}$ does not exist}  
    &\Longrightarrow \arg \min_{\vgamma_{[\alpha:q]}} f (\vgamma_{[\alpha:q]} \mid \vb_{[\alpha:q]}, \vlambda_{[\alpha:q]}) \in \cR_{q-\alpha+1, 1}^{m_\alpha + \cdots + m_q}
\end{dcases}.
\end{align*}
Thus, the condition such that $\vs_\alpha$ exists is given by
\begin{align*}
  \forall j \in \{ \alpha, \alpha + 1, \ldots, q \},\ d_{\alpha, j} > 0,
\end{align*}
where $d_{\alpha, j}$ is given in \eqref{d2}.
If $\vs_\alpha$ does not exist, we check the existence of $\vs_{\alpha+1}$, which is the minimizer of $f (\vgamma_{[\alpha+1:q]} \mid \vb_{[\alpha+1:q]}, \vlambda_{[\alpha+1:q]})$, by the same procedure.
By repeating the procedure for $\alpha = 1, 2, \ldots$, we have
\begin{align*}
  &\begin{dcases}
    \text{$\vs_1 \in \cB_q^{m_1 + \cdots + m_q}$ exists}
      &\Longrightarrow \vgamma^\ast = \vs_1 \\
    \text{$\vs_1 \in \cB_q^{m_1 + \cdots + m_q}$ does not exist}  
      &\Longrightarrow \vgamma^\ast \in \cR_{q, 1}^{m_1 + \cdots + m_q}
  \end{dcases}, \\
  &\vgamma^\ast \in \cR_{q, \alpha-1}^{m_1 + \cdots + m_q} \text{ and } \begin{dcases}
      \text{$\vs_\alpha \in \cB_{q - \alpha + 1}^{m_\alpha + \cdots m_q}$ exists}
        &\Longrightarrow \vgamma^\ast = \left( \0_{m_1 + \cdots + m_{\alpha-1}}', \vs_\alpha' \right)' \\
      \text{$\vs_\alpha \in \cB_{q - \alpha + 1}^{m_\alpha + \cdots m_q}$ does not exist}
        &\Longrightarrow \vgamma^\ast \in \cR_{q, \alpha}^{m_1 + \cdots m_q}
    \end{dcases}.
\end{align*}
Consequently, Theorems~\ref{th main} and \ref{th main2} are proved.

%=============================
% Appendix A.3
%=============================
\subsection{Proof of Lemma~\ref{lem1}}
\label{ap lem1}

Since $\vgamma_1 \neq \0_{m_1} \Leftrightarrow \| \vgamma_{(j)} \| \neq 0$ holds for all $j \in \{1, \ldots, q\}$, \eqref{s-eq} exists.
We first derive $\vgamma$ satisfying \eqref{s-eq}.
When $j = 1$, from $\| \vgamma_{(1)} \| = \| \vgamma_1 \|$, \eqref{s-eq} implies
\begin{align*}
  \left( 1 + \sum_{\ell=2}^q \dfrac{\lambda_\ell}{\| \vgamma_{(\ell)} \|} + \dfrac{\lambda_1}{\| \vgamma_1 \|} \right) \| \vgamma_1 \| = \| \vb_1 \|
    \Longleftrightarrow \left( 1 + \sum_{\ell=2}^q \dfrac{\lambda_\ell}{\| \vgamma_{(\ell)} \|} \right) \| \vgamma_1 \| = d_1 > 0.
\end{align*}
When $j = j_0\ (\in \{ 1, \ldots, q-2 \})$, we suppose 
\begin{align}
\label{MIas}
  \left( 1 + \sum_{\ell=j_0+1}^q \dfrac{\lambda_\ell}{\| \vgamma_{(\ell)} \|} \right) \| \vgamma_{(j_0)} \| = d_{j_0} > 0.
\end{align}
Notice that $\| \vgamma_{j+1} \|^2 + \| \vgamma_{(j)} \|^2 = \| \vgamma_{(j+1)} \|^2$. 
Hence, squaring \eqref{s-eq} for $j = j_0 + 1$ and \eqref{MIas} and adding the results yields 
\begin{align*}
  \left( 1 + \sum_{\ell=j_0+1}^q \dfrac{\lambda_\ell}{\| \vgamma_{(\ell)} \|} \right)^2 \| \vgamma_{(j_0+1)} \|^2 = d_{j_0}^2 + \| \vb_{j_0+1} \|^2.
\end{align*}
By taking the square root of both sides in the above equation, it follows from the same procedure when $j = 1$ that
\begin{align*}
  &\left( 1 + \sum_{\ell=j_0+1}^q \dfrac{\lambda_\ell}{\| \vgamma_{(\ell)} \|} \right) \| \vgamma_{(j_0 + 1)} \| 
    = \left( 1 + \sum_{\ell=j_0+2}^q \dfrac{\lambda_\ell}{\| \vgamma_{(\ell)} \|} \right) \| \vgamma_{(j_0 + 1)} \| 
         + \lambda_{j_0+1}   
    = \sqrt{d_{j_0}^2 + \| \vb_{j_0+1} \|^2} \\
  &\qquad \quad \Longleftrightarrow
    \left( 1 + \sum_{\ell=j_0+2}^q \dfrac{\lambda_\ell}{\| \vgamma_{(\ell)} \|} \right) \| \vgamma_{(j_0 + 1)} \| 
      = d_{j_0+1} > 0.
\end{align*}
Hence, mathematical induction gives 
\begin{align}
\label{MIres}
  \left( 1 + \sum_{\ell=j+1}^q \dfrac{\lambda_\ell}{\| \vgamma_{(\ell)} \|} \right) \| \vgamma_{(j)} \| = d_j > 0,
\end{align}
for $j = 1, \ldots, q-1$.
Furthermore, the same procedure with \eqref{s-eq} for $j = q$ and \eqref{MIas} for $j_0 = q-1$ yields
\begin{align*}
  \left( 1 + \dfrac{\lambda_q}{\| \vgamma \|} \right) \| \vgamma \| = \sqrt{d_{q-1}^2 + \| \vb_q \|^2}.
\end{align*}
From the results above, we have
\begin{align*}
  \| \vgamma_{(j)} \| = \begin{dcases}
      \left( 1 + \sum_{\ell=j+1}^q \dfrac{\lambda_\ell}{\| \vgamma_{(\ell)} \|} \right)^{-1} d_j
        & (j = 1, \ldots, q-1) \\
       d_q  & (j = q)
    \end{dcases}.
\end{align*}
Furthermore, this result implies 
\begin{align*}
  \| \vgamma_{(j)} \|
    &= \left( 1 + \sum_{\ell=j+1}^q \dfrac{\lambda_\ell}{\| \vgamma_{(\ell)} \|} \right)^{-1} d_j
    = \left( 1 + \sum_{\ell=j+2}^q \dfrac{\lambda_\ell}{\| \vgamma_{(\ell)} \|} + \dfrac{\lambda_{j+1}}{\| \vgamma_{(j+1)} \|} \right)^{-1} d_j, \\
  \left( 1 + \sum_{\ell=j+1}^q \dfrac{\lambda_\ell}{\| \vgamma_{(\ell)} \|} \right) 
    &= \dfrac{d_j}{\| \vgamma_{(j)} \|},
\end{align*}
for $j = 1, \ldots, q-2$ and $\| \vgamma_{(q)} \| = d_q$.
Hence, for $j = 1, \ldots, q-1$, we have
\begin{align*}
  \| \vgamma_{(j)} \|
    &= \left( \dfrac{d_{j+1}}{\| \vgamma_{(j+1)} \|} + \dfrac{\lambda_{j+1}}{\| \vgamma_{(j+1)} \|} \right)^{-1} d_j
    = \dfrac{d_j}{d_{j+1} + \lambda_{j+1}} \| \vgamma_{(j+1)} \|.
\end{align*}
Since $\| \vgamma_{[1:q]} \| = d_q$, the above equation implies
\begin{align*}
  \dfrac{\| \vgamma_{(j)} \|}{d_j}
    &= \dfrac{d_{j+1}}{d_{j+1} + \lambda_{j+1}} \dfrac{\| \vgamma_{(j+1)} \|}{d_{j+1}}
    = \dfrac{d_{j+1}}{d_{j+1} + \lambda_{j+1}} \dfrac{d_{j+2}}{d_{j+2} + \lambda_{j+2}} \dfrac{\| \vgamma_{[1:j+2]} \|}{d_{j+2}} \\
    &= \cdots
    = \prod_{\ell=j+1}^q \dfrac{d_\ell}{d_\ell + \lambda_\ell} \dfrac{\| \vgamma_{[1:q]} \|}{d_q}
    = \prod_{\ell=j+1}^q \dfrac{d_\ell}{d_\ell + \lambda_\ell}.
\end{align*}
Hence, we have
\begin{align*}
  \| \vgamma_1 \|
    &= d_1 \prod_{\ell=2}^q \dfrac{d_\ell}{d_\ell + \lambda_\ell}
    = (d_1 + \lambda_1) \prod_{\ell=1}^q \dfrac{d_\ell}{d_\ell + \lambda_\ell}
    = \| \vb_1 \| \prod_{\ell=1}^q \dfrac{d_\ell}{d_\ell + \lambda_\ell}, \\
  \| \vgamma_j \|
    &= \sqrt{ \| \vgamma_{(j)} \|^2 - \| \vgamma_{(j-1)} \|^2}
    = \sqrt{ \| \vgamma_{(j)} \|^2 - \left( \dfrac{d_{j-1}}{d_j + \lambda_j} \right)^2 \| \vgamma_{(j)} \|^2}
    = \dfrac{\| \vb_j \|}{d_j + \lambda_j} \| \vgamma_{(j)} \| \\
    &= \| \vb_j \| \dfrac{d_j}{d_j + \lambda_j} \dfrac{\| \vgamma_{(j)} \|}{d_j}
    = \| \vb_j \| \prod_{\ell=j}^q \dfrac{d_\ell}{d_\ell + \lambda_\ell} \quad (j = 2, \ldots, q).
\end{align*}
Moreover, it follows from \eqref{s-eq} that
\begin{align*}
  \| \vgamma_j \| 
    = \left( 1 + \sum_{\ell=j}^q \dfrac{\lambda_\ell}{\| \vgamma_{(\ell)} \|} \right)^{-1} \| \vb_j \|,
\end{align*}
and we have
\begin{align*}
  \left( 1 + \sum_{\ell=j}^q \dfrac{\lambda_\ell}{\| \vgamma_{(\ell)} \|} \right)^{-1}
    = \prod_{\ell=j}^q \dfrac{d_\ell}{d_\ell + \lambda_\ell}.
\end{align*}
Thus, $\vgamma$ satisfying \eqref{s-eq} is given by
\begin{align*}
  \vgamma_j
    = \prod_{\ell=j}^q \dfrac{d_\ell}{d_\ell + \lambda_\ell} \vb_j \quad (j = 1, \ldots, q).
\end{align*}

Next, we give a necessary and sufficient condition for a root of \eqref{s-eq} existing.
As described above, we have the following sufficient condition:
\begin{align*}
  \forall j \in \{ 1, \ldots, q \},\ d_j > 0
    \Longrightarrow \exists \vgamma \in \bR^{m_1 + \cdots + m_q}\ s.t.\ \text{\eqref{s-eq} holds}.
\end{align*}
On the other hand, if $j$ exists such that $d_j \le 0$ holds, \eqref{MIres} does not hold, and hence a real root of \eqref{s-eq} does not exist.
This implies 
\begin{align*}
  \exists j \in \{ 1, \ldots, q \}\ s.t.\ d_j \le 0
    \Longrightarrow \forall \vgamma \in \bR^{m_1 + \cdots + m_q},\ \text{\eqref{s-eq} does not hold}.
\end{align*}
The contrapositive of this leads to the following necessary condition:
\begin{align*}
  \exists \vgamma \in \bR^{m_1 + \cdots + m_q}\ s.t.\ \text{\eqref{s-eq} holds}
    \Longrightarrow \forall j \in \{ 1, \ldots, q \},\ d_j > 0.
\end{align*}
Thus, we have the necessary and sufficient condition. Consequently, Lemma~\ref{lem1} is proved.

%=============================
% Appendix A.4
%=============================
\subsection{Proof of Theorem~\ref{th lam}}
\label{ap thlam}

We set $\lambda_j = \lambda w_j / L$ for $f$ in \eqref{f}.
Theorem~\ref{th main} implies a sufficient condition such that $f$ is minimized at $\vgamma = \0_q$ as
\begin{align}
\label{lamSC}
  &\forall j \in \{ 1, \ldots, q \}, d_{j, j} = 0 
  \Longleftrightarrow 
    \lambda \ge \max_{j \in \{ 1, \ldots, q \}} \dfrac{L |b_j|}{w_j}.
\end{align}
With this condition, we derive the condition for $\lambda$ such that $\PRSS (\vtheta \mid \delta, \lambda)$ is minimized at $\vtheta = \0_{kq}$. 
Notice that the optimality of Algorithm \ref{alg main1} means that we can obtain the optimal solution for an arbitrary initial solution.
Hence, it is sufficient to consider the condition such that $\PRSS^+ (\vtheta \mid \0_{kq}, \delta, \lambda)$ is minimized at $\vtheta = \0_{kq}$.
At the minimization of \eqref{theta ell}, a sufficient condition for $\lambda$ such that \eqref{theta ell} is minimized at $\vtheta_\ell = \0_q$ when $\hat{\vtheta} = \0_{kq}$ follows \eqref{lamSC}.
Consequently, Theorem~\ref{th lam} is proved.

%=============================
% Appendix A.5
%=============================
\subsection{Details of the transformation of the matrix of basis functions}
\label{ap QR}

Let $\vF_q$ be a $q \times q$ anti-diagonal matrix with anti-diagonal elements ones.
Then, the QR decomposition yields $(\vv_q, \vv_{q-1}, \ldots, \vv_1) = \vV \vF_q = \vH_0 \vQ_0$, where $\vH_0$ is a $p \times q$ matrix satisfying $\vH_0' \vH_0 = \vI_q$ and $\vQ_0$ is a $q \times q$ upper triangular matrix.
From $\vF_q \vF_q = \vI_q$, this decomposition implies $\vV = \vH_0 \vF_q \vF_q \vQ_0 \vF_q$ and we have $\vH = \vH_0 \vF_q$ and $\vQ = \vF_q \vQ_0 \vF_q$.

%=============================
% Appendix A.6
%=============================
\subsection{Simulation results for $\mathrm{SNR} = 3$}
\label{ap SNR3}

This section shows the simulation results described in Section~\ref{sec sim} when $\mathrm{SNR} = 3$.
In comparison with $\mathrm{SNR} = 1$, we can see that all the methods performed better.

\begin{table}[h]
\caption{\ Selection probability (\%) when $q = 6$ \label{tab SP1snr3}}
\centering
\begin{tabular}{rrrrrrrrrrr}
\toprule
  &  &  & \multicolumn{4}{c}{Variable} & \multicolumn{4}{c}{Degree} \\
\cmidrule(lr){4-7} \cmidrule(lr){8-11}
 $n$ & $p$ & $k$ & HOGL1 & HOGL2 & SGL & wSCAD & HOGL1 & HOGL2 & SGL & wSCAD \\
\midrule
 100 & 10 & 10 & 28.2 & \bf 91.6 & 19.5 & 6.9 & 12.9 & \bf 72.7 & 4.7 & 0.9 \\
 300 & 10 & 10 & 70.8 & \bf 97.7 & 55.9 & 85.4 & 40.8 & \bf 92.4 & 14.2 & 75.2 \\
 500 & 10 & 10 & 82.9 & \bf 98.2 & 69.5 & 95.2 & 52.6 & \bf 95.1 & 23.0 & 85.2 \\ \midrule
 100 & 40 & 10 & 9.5 & \bf 70.1 & 9.5 & 4.4 & 8.1 & \bf 54.9 & 2.0 & 0.1 \\
 300 & 120 & 10 & 64.7 & 89.1 & 59.2 & \bf 91.2 & 36.3 & 82.3 & 7.4 & \bf 83.5 \\
 500 & 200 & 10 & 84.0 & 93.1 & 81.6 & \bf 98.8 & 51.7 & 85.6 & 15.8 & \bf 96.2 \\ \midrule
 100 & 10 & 40 & 17.8 & \bf 82.7 & 6.0 & 0.1 & 7.1 & \bf 68.9 & 5.5 & 0.1 \\
 300 & 10 & 120 & 29.0 & \bf 94.6 & 12.5 & 16.5 & 16.8 & \bf 89.7 & 10.3 & 83.4 \\
 500 & 10 & 200 & 62.8 & \bf 97.5 & 21.7 & 14.3 & 22.2 & \bf 94.2 & 6.7 & 92.7 \\ \midrule
 100 & 20 & 40 & 7.3 & \bf 69.7 & 2.0 & 0.5 & 5.3 & \bf 63.2 & 4.0 & 0.0 \\
 300 & 60 & 120 & 36.4 & \bf 91.5 & 11.2 & 82.6 & 9.9 & 86.2 & 4.0 & \bf 93.9 \\
 500 & 100 & 200 & 76.8 & \bf 92.5 & 34.8 & 87.9 & 26.1 & 93.0 & 5.6 & \bf 97.3 \\
\bottomrule
\end{tabular}
\end{table}
\begin{table}[h]
\caption{\ MSE when $q = 6$ \label{tab MSE1snr3}}
\centering
\begin{tabular}{rrrrrrrrrrr}
\toprule
  &  &  & \multicolumn{4}{c}{$\MSE_\mathrm{f}$} & \multicolumn{4}{c}{$\MSE_\mathrm{c}$} \\
\cmidrule(lr){4-7} \cmidrule(lr){8-11}
 $n$ & $p$ & $k$ & HOGL1 & HOGL2 & SGL & wSCAD & HOGL1 & HOGL2 & SGL & wSCAD \\
\midrule
 100 & 10 & 10 & 0.061 & \bf 0.037 & 0.064 & 0.086 & 0.438 & \bf 0.176 & 0.619 & 1.273 \\
 300 & 10 & 10 & 0.017 & \bf 0.011 & 0.020 & 0.035 & 0.096 & \bf 0.044 & 0.171 & 0.248 \\
 500 & 10 & 10 & 0.009 & \bf 0.006 & 0.011 & 0.013 & 0.048 & \bf 0.023 & 0.095 & 0.091 \\ \midrule
 100 & 40 & 10 & 0.027 & 0.020 & 0.030 & \bf 0.018 & 1.317 & \bf 0.644 & 2.828 & 3.623 \\
 300 & 120 & 10 & 0.005 & 0.004 & 0.005 & \bf 0.001 & 0.409 & \bf 0.184 & 1.055 & 0.462 \\
 500 & 200 & 10 & 0.003 & 0.002 & 0.003 & \bf 0.000 & 0.247 & \bf 0.113 & 0.632 & 0.131 \\ \midrule
 100 & 10 & 40 & 0.197 & \bf 0.046 & 0.147 & 0.622 & 0.469 & \bf 0.060 & 0.222 & 0.439 \\
 300 & 10 & 120 & 0.061 & \bf 0.012 & 0.059 & 0.126 & 0.042 & \bf 0.004 & 0.032 & 0.052 \\
 500 & 10 & 200 & 0.031 & \bf 0.007 & 0.035 & 0.036 & 0.011 & \bf 0.001 & 0.012 & 0.010 \\ \midrule
 100 & 20 & 40 & 0.118 & \bf 0.032 & 0.087 & 0.219 & 0.812 & \bf 0.121 & 0.389 & 0.805 \\
 300 & 60 & 120 & 0.016 & \bf 0.005 & 0.015 & 0.030 & 0.105 & \bf 0.014 & 0.087 & 0.127 \\
 500 & 100 & 200 & 0.005 & \bf 0.003 & 0.005 & 0.003 & 0.025 & \bf 0.005 & 0.030 & 0.017 \\
\bottomrule
\end{tabular}
\end{table}
\begin{table}[h]
\caption{\ Selection probability (\%) when $q = 10$ \label{tab SP2snr3}}
\centering
\begin{tabular}{rrrrrrrrrrr}
\toprule
  &  &  & \multicolumn{4}{c}{Variable} & \multicolumn{4}{c}{Degree} \\
\cmidrule(lr){4-7} \cmidrule(lr){8-11}
 $n$ & $p$ & $k$ & HOGL1 & HOGL2 & SGL & wSCAD & HOGL1 & HOGL2 & SGL & wSCAD \\
\midrule
 100 & 10 & 10 & 0.8 & \bf 88.0 & 18.3 & 0.4 & 0.0 & \bf 57.6 & 0.0 & 0.0 \\
 300 & 10 & 10 & 0.4 & \bf 97.6 & 32.8 & 0.4 & 0.0 & \bf 88.3 & 0.0 & 0.0 \\
 500 & 10 & 10 & 0.1 & \bf 98.1 & 38.2 & 0.3 & 0.0 & \bf 92.7 & 0.0 & 0.0 \\ \midrule
 100 & 40 & 10 & 0.4 & \bf 67.6 & 22.4 & 0.6 & 0.0 & \bf 40.1 & 0.0 & 0.0 \\
 300 & 120 & 10 & 0.1 & \bf 89.5 & 47.1 & 0.9 & 0.0 & \bf 78.0 & 0.0 & 0.1 \\
 500 & 200 & 10 & 0.0 & \bf 93.5 & 60.5 & 1.8 & 0.0 & \bf 84.0 & 0.0 & 0.2 \\ \midrule
 100 & 10 & 40 & 0.4 & \bf 75.3 & 13.8 & 0.0 & 0.0 & \bf 50.6 & 0.0 & 0.0 \\
 300 & 10 & 120 & 0.3 & \bf 92.8 & 9.0 & 0.0 & 0.0 & \bf 81.5 & 0.0 & 0.0 \\
 500 & 10 & 200 & 0.1 & \bf 95.7 & 7.4 & 0.0 & 0.0 & \bf 90.1 & 0.0 & 0.0 \\ \midrule
 100 & 20 & 40 & 0.3 & \bf 65.0 & 5.7 & 0.0 & 0.0 & \bf 46.9 & 0.0 & 0.0 \\
 300 & 60 & 120 & 0.0 & \bf 90.1 & 13.6 & 1.5 & 0.0 & \bf 79.4 & 0.0 & 0.0 \\
 500 & 100 & 200 & 0.0 & \bf 92.7 & 17.1 & 0.3 & 0.0 & \bf 90.3 & 0.0 & 0.0 \\
\bottomrule
\end{tabular}
\end{table}
\begin{table}[h]
\caption{\ MSE when $q = 10$ \label{tab MSE2snr3}}
\centering
\begin{tabular}{rrrrrrrrrrr}
\toprule
  &  &  & \multicolumn{4}{c}{$\MSE_\mathrm{f}$} & \multicolumn{4}{c}{$\MSE_\mathrm{c}$} \\
\cmidrule(lr){4-7} \cmidrule(lr){8-11}
 $n$ & $p$ & $k$ & HOGL1 & HOGL2 & SGL & wSCAD & HOGL1 & HOGL2 & SGL & wSCAD \\
\midrule
 100 & 10 & 10 & 0.127 & \bf 0.040 & 0.162 & 0.266 & 4276.068 & \bf 0.826 & 4004.717 & 4310.401 \\
 300 & 10 & 10 & 0.041 & \bf 0.011 & 0.049 & 0.065 & 1262.396 & \bf 0.034 & 1129.817 & 1271.469 \\
 500 & 10 & 10 & 0.025 & \bf 0.007 & 0.029 & 0.030 & 741.832 & \bf 0.018 & 661.475 & 751.059 \\ \midrule
 100 & 40 & 10 & 0.049 & \bf 0.021 & 0.048 & 0.048 & 1737.444 & \bf 1.019 & 1579.390 & 1339.786 \\
 300 & 120 & 10 & 0.008 & 0.004 & 0.007 & \bf 0.004 & 861.548 & \bf 0.266 & 739.860 & 632.382 \\
 500 & 200 & 10 & 0.004 & 0.002 & 0.003 & \bf 0.001 & 577.077 & \bf 0.101 & 483.212 & 427.223 \\ \midrule
 100 & 10 & 40 & 2.155 & \bf 0.052 & 0.915 & 2.352 & 31.219 & \bf 0.071 & 1056.697 & 98.678 \\
 300 & 10 & 120 & 1.431 & \bf 0.013 & 0.550 & 2.309 & 6.919 & \bf 0.004 & 133.269 & 7.512 \\
 500 & 10 & 200 & 1.022 & \bf 0.007 & 0.410 & 2.011 & 2.401 & \bf 0.001 & 49.203 & 3.439 \\ \midrule
 100 & 20 & 40 & 0.648 & \bf 0.035 & 0.339 & 1.028 & 20.777 & \bf 0.113 & 259.949 & 148.508 \\
 300 & 60 & 120 & 0.163 & \bf 0.005 & 0.064 & 0.099 & 6.517 & \bf 0.011 & 65.318 & 94.937 \\
 500 & 100 & 200 & 0.083 & \bf 0.003 & 0.030 & 0.045 & 3.109 & \bf 0.004 & 28.790 & 48.901 \\
\bottomrule
\end{tabular}
\end{table}
\begin{table}[h]
\caption{\ Runtime (sec.) \label{tab timesnr3}}
\centering
\begin{tabular}{rrrrrrrrrrr}
\toprule
  &  &  & \multicolumn{4}{c}{$q=6$} & \multicolumn{4}{c}{$q=10$} \\
\cmidrule(lr){4-7} \cmidrule(lr){8-11}
 $n$ & $p$ & $k$ & HOGL1 & HOGL2 & SGL & wSCAD & HOGL1 & HOGL2 & SGL & wSCAD \\
\midrule
 100 & 10 & 10 & 138.8 & \bf 4.9 & 10.9 & 17.6 & 148.3 & \bf 7.0 & 7.2 & 20.0 \\
 300 & 10 & 10 & 112.2 & \bf 4.3 & 10.4 & 15.8 & 167.4 & \bf 6.2 & 7.7 & 19.3 \\
 500 & 10 & 10 & 94.5 & \bf 4.1 & 10.7 & 14.3 & 175.3 & \bf 5.9 & 8.2 & 20.0 \\ \midrule
 100 & 40 & 10 & 137.2 & \bf 4.9 & 9.9 & 16.4 & 168.2 & 6.8 & \bf 6.8 & 19.7 \\
 300 & 120 & 10 & 95.3 & \bf 4.4 & 16.8 & 14.4 & 189.8 & \bf 5.8 & 23.7 & 18.4 \\
 500 & 200 & 10 & 77.9 & \bf 5.8 & 43.5 & 16.1 & 194.1 & \bf 7.3 & 66.9 & 26.2 \\ \midrule
 100 & 10 & 40 & 1203.6 & \bf 33.0 & 133.0 & 76.8 & 1678.3 & \bf 45.8 & 83.4 & 94.3 \\
 300 & 10 & 120 & 3533.1 & \bf 85.9 & 872.9 & 455.0 & 14674.1 & \bf 122.9 & 680.3 & 1090.2 \\
 500 & 10 & 200 & 4775.6 & \bf 132.1 & 2538.0 & 1561.1 & 42591.9 & \bf 193.5 & 2091.8 & 4737.2 \\ \midrule
 100 & 20 & 40 & 1183.6 & \bf 30.6 & 93.0 & 72.0 & 1954.3 & \bf 43.1 & 44.5 & 83.0 \\
 300 & 60 & 120 & 3011.6 & \bf 62.2 & 936.3 & 412.1 & 17813.8 & \bf 104.8 & 534.7 & 1044.1 \\
 500 & 100 & 200 & 2874.1 & \bf 87.4 & 3410.7 & 1277.9 & 47408.6 & \bf 125.6 & 2769.9 & 4189.2 \\
\bottomrule
\end{tabular}
\end{table}

\end{document}